\documentclass[longauth]{aa}  

\usepackage[colorlinks]{hyperref}

\usepackage{booktabs}
\usepackage{tabularx}
\usepackage{scrextend}
\usepackage{tablefootnote}
\let\orgautoref\autoref
\renewcommand{\autoref}
        {\def\equationautorefname{Eq.}%
         \def\figureautorefname{Fig.}%
         \def\sectionautorefname{Sect.}%
         \def\subsectionautorefname{Sect.}%
         \def\subsubsectionautorefname{Sect.}%
         \orgautoref}

\usepackage{xcolor}
\definecolor{dark-red}{rgb}{0.9,0.0,0.0}
\definecolor{dark-blue}{rgb}{0.15,0.15,0.9}
\definecolor{dark-green}{rgb}{0.15,0.8,0.15}
\definecolor{medium-blue}{rgb}{0,0,0.9}
\hypersetup{linkcolor={dark-blue},citecolor={dark-blue}, urlcolor={medium-blue}
}

\newcommand{\topline}{
    \hline\hline
    \noalign{\smallskip}
}
\newcommand{\midline}{
      \noalign{\smallskip}
      \hline
      \noalign{\smallskip}
}
\newcommand{\bottomline}{
    \noalign{\smallskip}
    \hline
}

\makeatletter
\renewcommand*\aa@pageof{, page \thepage{} of \pageref*{LastPage}} 
\makeatother

\usepackage{changepage} 
\usepackage{adjustbox}
\usepackage{rotating}

\usepackage{graphicx}
\usepackage{txfonts}
\usepackage{soul}
\usepackage[
    separate-uncertainty=true,
    multi-part-units=single,
    range-phrase=--,
    range-units=single
]{siunitx}
\ifdefined\unit\else
  \ifdefined\NewCommandCopy
    \NewCommandCopy\unit\si
  \else
    \NewDocumentCommand\unit{O{}m}{\si[#1]{#2}}
  \fi
\fi
\DeclareSIUnit{\ppm}{ppm}
\DeclareSIUnit{\year}{yr}
\DeclareSIUnit{\julianday}{JD}
\DeclareSIUnit{\au}{au}
\DeclareSIUnit{\solarmass}{M\textsubscript{\(\odot\)}}
\DeclareSIUnit{\solarradius}{R\textsubscript{\(\odot\)}}
\DeclareSIUnit{\solarluminosity}{L\textsubscript{\(\odot\)}}
\DeclareSIUnit{\earthmass}{M\textsubscript{\(\oplus\)}}
\DeclareSIUnit{\jupitermass}{M\textsubscript{J}}
\DeclareSIUnit{\earthflux}{\text{$\Phi$}\textsubscript{\(\oplus\)}}
\defcitealias{Stefanov2025a}{S25}
\defcitealias{Lovis2005}{L05}

\begin{document} 

   \title{Long-period magnetic activity in the K dwarf GJ\,1137 and a new super-Earth on a 9-day orbit}

    \subtitle{A cautionary tale for Jovian-analogue detection from Doppler surveys\thanks{
    Based on observations collected at 
the European Organization for Astronomical Research in the Southern Hemisphere under ESO 
programmes 072.C-0488,
183.C-0972,
091.C-0936,
192.C-0852,
198.C-0836}}
  
\author{
Denitza\,Stoeva\inst{1} \and
Atanas\,K.\,Stefanov\inst{2,3} \and 
Stefan\,Y.\,Stefanov\inst{1,4} \and
Marina\,Lafarga\inst{6,7} \and
Elena\,Vchkova Bebekovska\inst{5} \and 
Simone\,Filomeno\inst{8,9,10} \and
Jonay\,I.\,González\,Hernández\inst{2,3} \and
Alejandro\,Suárez\,Mascareño\inst{2,3} \and
Rafael\,Rebolo\inst{2,3,11} \and
Nicola\,Nari\inst{2,3,12} \and
Júlia\,M.\,Mestre\inst{13} \and
Desislava\,Antonova\inst{1} \and
Evelina\,Zaharieva\inst{1} \and
Vladimir\,Bozhilov\inst{1} \and
Trifon\,Trifonov\inst{1,14} 
}

    \institute{Department of Astronomy, Sofia University "St Kliment Ohridski", 5 James Bourchier Blvd, BG-1164 Sofia, Bulgaria
    \and
    Instituto de Astrofísica de Canarias, 38205 La Laguna, Spain
    \and
    Departamento de Astrofísica, Universidad de La Laguna, 38206 La Laguna, Spain
    \and 
    Institute of Astronomy and National Astronomical Observatory, Bulgarian Academy of Sciences, 72 Tsarigradsko shosse Blvd., 1784 Sofia, Bulgaria 
    \and
        Ss. Cyril and Methodius University in Skopje, Faculty of Natural Sciences and Mathematics-Skopje, Institute of Physics, Arhimedova 3 1000 Skopje, North Macedonia
    \and 
    Department of Physics, University of Warwick, Coventry, CV4 7AL, UK
    \and
Centre for Exoplanets and Habitability, University of Warwick, Coventry, CV4 7AL, UK
   \and 
   INAF-Osservatorio Astronomico di Roma, Via Frascati 33, I-00040 Monte Porzio Catone (RM), Italy
   \and
Dipartimento di Fisica, Universit\`a di Roma Tor Vergata, Via della Ricerca Scientifica 1, I-00133 Roma, Italy
\and
Dipartimento di Fisica, Sapienza Università di Roma, Piazzale Aldo Moro 5, 00185 Roma, Italy
\and
Consejo Superior de Investigaciones Cient\'ificas (CSIC), 28006 Madrid, Spain
\and
Light Bridges S. L., 35004 Las Palmas de Gran Canaria, Spain
\and
Dipartimento di Fisica e Astronomia "Galileo Galilei", Universit\`a di Padova, Vicolo dell’Osservatorio 3, 35122 Padova, Italy
\and
Landessternwarte, Zentrum f\"ur Astronomie der Universt\"at Heidelberg, K\"onigstuhl 12, 69117 Heidelberg, Germany\\}

   \date{Received Xxxxx xx, 2025; accepted Xxxxx xx, 2025}

  \abstract
{Detection and characterisation of Jovian analogues in precision radial velocity (RV) measurements is gaining momentum due to the constantly increasing observational baseline of Doppler surveys. The occurrence rate of Jovian-mass exoplanets is crucial to understanding the architecture of planetary systems. 
However, long-period RV signals in Doppler surveys could also be induced by stellar magnetic cycles, 
leading to misinterpretations of planetary candidates.}
{We investigate long-term RV variability in the K-dwarf star GJ\,1137 (HD\,93083, HIP\,52521), a known Saturn-mass
exoplanet host, and assess the role of stellar activity in shaping the observed signals.}
{We analyse 13 years of archival high-precision spectroscopic observations obtained with the High Accuracy Radial velocity Planet Searcher spectrograph (HARPS). We performed an extensive spectroscopic analysis of the stellar activity indicators
and applied an RV modelling approach, incorporating Keplerian fits, Gaussian process regression as a proxy for stellar activity, and other stellar activity diagnostics. Furthermore, we refined
the orbital parameters and the minimum mass of the known exoplanet GJ 1137\,b and searched for
additional planetary candidates in the system.}
{We detect a long-period RV signal that, if interpreted as planetary, would suggest the presence of a Jovian analogue companion. However, our spectroscopic activity analysis provides strong evidence that this variability is induced by the star’s long-term magnetic cycle \mbox{( $P_\text{cyc} = 5870^{+480}_{-350}$ days)}
rather than by an orbiting planet. The signal is detected in both full width at half maximum (FWHM) of the cross-correlation function and the chromospheric activity index \mbox{$\log R'_{\rm HK}$}.
We measure the stellar rotation period to $P_\text{rot}=32.3^{+1.2}_{-1.3}\,\unit\day$ and identify a significant short-period RV signal, which we attribute to a Super Earth with a period of $9.6412^{+(12)}_{-(11)}\,\unit\day$ and a 
minimum mass of \mbox{$5.12^{+0.70}_{-0.69}\,\unit\earthmass$}, making GJ\,1137 a multiple-planet system.}
{}

   \keywords{
methods: data analysis -- planets and satellites: detection -- stars: activity -- techniques: radial velocities         
               }

   \maketitle
%

\section{Introduction}
 
Over the past three decades, the field of exoplanet research has witnessed remarkable growth, with more than 6000 
confirmed exoplanets as of October 2025.\footnote{Up-to-date statistics available at \url{ https://exoplanetarchive.ipac.caltech.edu}.} Of these, approximately 1100 have been detected using the radial velocity (RV) technique, while the vast majority of more than 4400 confirmed planets have been identified via the transit method both from ground-based observations \citep{Pollacco2006, Bakos2004, Gillon2017}, and the unprecedented contributions of NASA’s {\it Kepler} space telescope \citep{Borucki2010} and the ongoing Transiting Exoplanet Survey Satellite \citep[TESS;][]{Ricker2015}. Additional discoveries stem from complementary techniques, including direct imaging of wide-orbit companions \citep[e.g.][]{Marois2008} and gravitational microlensing, which is particularly effective for detecting planets around distant and faint stars \citep[e.g.][]{Gaudi2012}.

Statistical analyses of large samples of surveyed stars indicate that planets are 
ubiquitous, with occurrence rates exceeding 0.5 planets per FGK-type star for orbital periods 
ranging from one day to several hundred days. These estimates are supported by Doppler surveys \citep{Howard2010} and transit surveys \citep{Petigura2013, Kunimoto2020}. Even higher occurrence rates have been found for planets orbiting M dwarfs, with estimates exceeding one planet per star \citep{Cassan2012, Bonfils2013, 
Dressing2015, Gaidos2016, Mignon2025}, and possibly increasing further from early- to mid-type M dwarfs \citep{Hardegree2019}.
The solar neighbourhood remains a prime hunting ground for exoplanets around low-mass stars, 
particularly M and K dwarfs, due to their abundance and relative proximity. Nearby stars are 
especially valuable for Doppler surveys and exoplanet characterisation because of their apparent brightness 
(resulting in higher signal-to-noise (S/N)observations). Many planet searches have therefore focused on them. Key efforts include the 
HARPS M-dwarf survey \citep{Bonfils2013, Astudillo-Defru2017a}, and several 
programmes targeting K dwarfs more specifically, such as the California Planet Survey \citep{Howard20102, Rosenthal2021}, the KOBE project \citep{Balsalobre-Ruza2025}, and the SOPHIE northern 
survey \citep{Hobson2018}. 
These surveys have significantly expanded the known exoplanet 
population in this stellar mass regime and continue to refine the occurrence rates and dynamical 
architectures of exoplanet systems around low-mass stars.
Nearby stars are also important targets for future observations with Extremely Large Telescopes (ELTs) 
such as ANDES at the ELT, Paranal, Chile \citep{Palle2025}, and are prime targets for direct-imaging campaigns
with NASA’s Habitable Worlds Observatory \citep{Harada2024} 
and ESA's LIFE interferometer \citep{Quanz2022}, owing to the greater angular separation between the star and any potential exoplanet.
 
In this paper, we present extensive orbital and stellar activity analysis of archival spectroscopic data of the K-dwarf star GJ\,1137 obtained with the High Accuracy Radial velocity Planet Searcher spectrograph \citep[HARPS,][]{Pepe2002,Mayor2003} mounted at the ESO 3.6\,m Telescope in La Silla, Chile. 
GJ\,1137 is a known exoplanet host, first reported by \citet{Lovis2005}, based on just 16 HARPS RVs over a baseline of approximately one year. This dataset helped us identify GJ\,1137\,b,  
a Saturn-mass exoplanet, with a minimum mass of $m_b \sin i = 0.37\,\unit\jupitermass$ and 
an orbital period of $P_b = 143.6$ days. The discovery paper also noted a long-term RV trend that suggested an additional massive companion. The public {\sc HARPS-RVB}ank spectral products released by \citet{Trifonov2020, Perdelwitz2024} now provide an excellent opportunity for a more detailed analysis of 
the orbital architecture of GJ\,1137\,b, and for a search for additional planets in the system.

In our work, we show evidence for a \SI{5640}{\day} RV signal in addition to GJ\,1137\,b. While this signal initially appears consistent with a Jovian-mass companion, we demonstrate that it is part of long-term stellar activity that we see in other spectroscopic activity indices. Furthermore, we perform full stellar activity and planet signal modelling, and report improved orbital and mass estimates for GJ\,1137\,b as well as stellar activity cycle properties for GJ\,1137, and its rotational period. Finally, we present significant evidence of a short-period exoplanet, GJ\,1137\,c, with a period of $9.64^{+0.12}_{-0.11}$ days and a 
minimum mass of $5.12^{+0.70}_{-0.69}$ \unit\earthmass.

The paper is organised as follows.
In \autoref{Sec2}, we introduce the available data applicable to our study. 
In \autoref{Sec3}, we present revised physical properties of the K-dwarf star GJ\,1137 by employing public data and state-of-the-art stellar parameter modelling.
Our orbital and spectroscopic data analysis scheme and results are introduced in \autoref{Sec4}, and finally, 
in \autoref{Sec5} we provide a summary and discussion of our findings.

\section{Data}
\label{Sec2}

 \subsection{HARPS data}

For our analysis of GJ\,1137, we rely on spectroscopic data obtained with the HARPS spectrograph. Our dataset comprises 140 high S/N spectra collected between January 2004 and July 2017. Of these, 87 spectra were obtained before and 53 after the optical fibre upgrade in May 2015 \citep{LoCurto2015}, hereafter referred to as HARPS-pre and HARPS-post, respectively. This upgrade improved the throughput of the HARPS instrument but changed the instrument profile. Thus, the pre- and post-fibre spectroscopic data products should be treated as if obtained from two separate spectrographs \citep[see][]{LoCurto2015, Trifonov2020}.

We use publicly available extracted RV and activity index data in the {\sc HARPS-RV} Bank. {\sc HARPS-RV} Bank contains precise RVs, as well as stellar activity indicators - $H\alpha, NaI \space  D, NaII \space D$, chromatic RV index $CRX$, and differential line width (dLW). All of these spectral products were derived via the {\sc SERVAL} pipeline \citep{Zechmeister2018}. In addition to  {\sc HARPS-RV} Bank data, we use the {\sc RACCOON} pipeline \citep{Lafarga2020} to compute the cross-correlation function (CCF) RVs \citep{Baranne1996} and derive other activity indicators. We first created a custom CCF mask using the high S/N template that SERVAL outputs, using default parameters for {\sc RACCOON} to select the mask lines. We then cross-correlated this mask with the observed spectra, following the methods explained in \citet{Lafarga2020}. In summary, {\sc RACCOON} computes a CCF order-by-order, coadds all the order CCFs together into a final CCF, and fits a Gaussian function to this final CCF to derive the spectrum RV from the CCF centroid. RACCOON also provides the standard CCF activity indicators: full width at half maximum (FWHM) and contrast, derived from the Gaussian fit to the CCF, and bisector inverse slope (BIS), calculated directly from the bisector of the CCF, as described by \citet{Queloz2001}. 

For orbital analysis, we used RVs taken from the {\sc HARPS-RV} Bank since these have been corrected for small, but significant RV systematics \citep[see][]{Trifonov2020}.
In addition, as shown by \citet{Escude2012} and \citet{Zechmeister2018}, for redder stars such as GJ\,1137 the co-adding stellar template method used by {\sc SERVAL} can deliver higher RV precision than the CCF method with a weighted binary mask used by {\sc RACCOON}. For studying the activity index data, we used the activity indices taken from both {\sc SERVAL} and {\sc RACCOON}.
All data used are listed in \autoref{tab:RVs}, the full table available in the CDS.

 \subsection{Photometry data}

TESS visited GJ\,1137 (TIC\,54237857) during sectors 3, 36, 63, and 90. The star has a TESS-band magnitude of $7.472\pm0.006$ mag \citep[TIC v8.2 catalogue; ][]{tessInputCatalogue}. In the available 120-second exposures, we expect it to produce a flux of $\sim 1.8 \times 10^7$ $\text{e}/\text{s}^{-1}$, which is nearly two orders of magnitude above the nominal saturation threshold.\footnote{See section 6.7.1 of the TESS Instrument Handbook and Data Release Notes.} As a result, the target exhibits pronounced blooming artefacts, rendering both the standard simple aperture photometry (SAP) and presearch data conditioning simple aperture photometry (PDC-SAP) light curves \citep[][]{Smith2012,Stumpe2012,Stumpe2014} unsuitable for analyses of long-term stellar variability. Visual inspection of the light curves reveals significant contamination that likely comes from charge spillover from saturated pixels, leading to a high incidence of low-quality data points. Nevertheless, we performed a transit search using the transit least-squares (TLS) periodogram algorithm by \citet{tls}. No statistically significant periodic signals were detected.

\begin{figure}
    \centering
    \resizebox{\hsize}{!}{\includegraphics{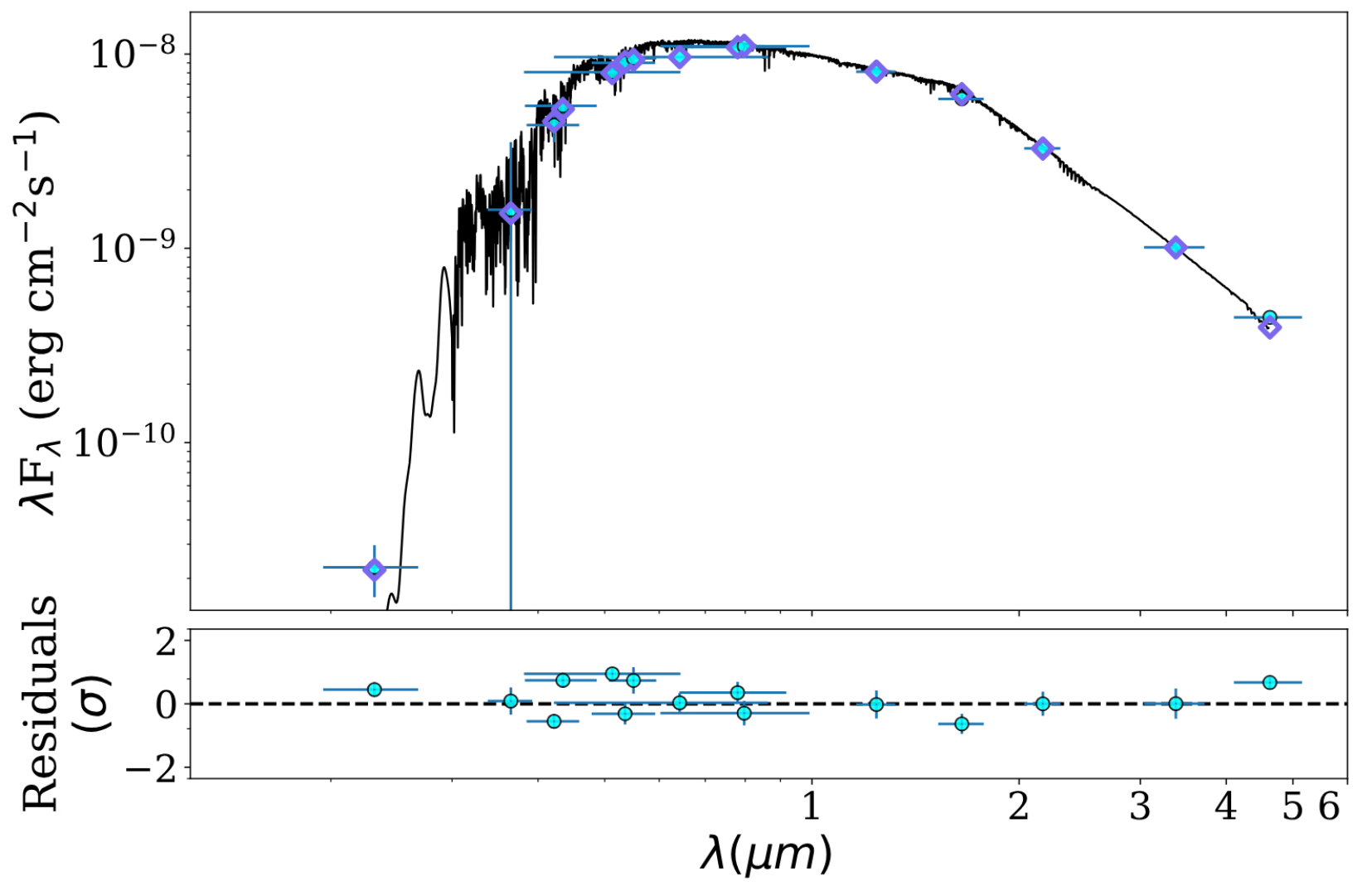}}
    \caption{
    SED fit for GJ\,1137. The best-fitting model, BT-Settl, is displayed in black, with normalised residuals shown below. Blue points represent flux values from photometry, while purple diamonds show the flux from synthetic photometry in the same passband.
    }
    \label{fig:SED}
\end{figure}

We attempted to determine the rotational period of this bright star, using photometric data from the All-Sky Automated Survey for Supernovae \citep[ASAS-SN; ][]{asassn1, asassn2}. Recently, the ASAS-SN reduction pipeline included a specialised module tailored for highly saturated stars \citep{Winecki&Kochanek2024}. Data processed using this approach were obtained through the publicly available \mbox{ASAS-SN Sky Patrol v1.0}\footnote{\url{https://asas-sn.osu.edu/}} by choosing the `Saturated stars' option. For consistency, we used only the dataset in $g$~band from observing seasons 2018 to 2025. Due to the proximity of GJ\,1137 to the ecliptic, some of the photometric measurements are affected by lunar contamination. To mitigate this, we calculated the angular distance between the Moon and GJ\,1137 for each \mbox{ASAS-SN} measurement and retained only those in which the Moon is more than \SI{90}{\deg} away on the sky. After this, the selected data was analysed using the generalised Lomb-Scargle algorithm \citep[GLS; ][]{Lomb1976, Scargle1982,Zechmeister2009}. We found no significant peaks in this data sample; see \autoref{fig:assasn_timeseries}.

\begin{table}[htp]

\renewcommand{\arraystretch}{1.2}
\caption{Stellar parameters of GJ\,1137 and their 1$\sigma$ uncertainties.} 
\label{table:phys_param}    

\centering          
\begin{tabular}{ l l l@{} c@{}}     
\hline\hline  \noalign{\vskip 0.5mm}        
  Parameter   & \citet{Lovis2005} & This work  & Ref. \\  
\hline    \noalign{\vskip 0.5mm}                   
   Spectral type                           &K3 V  & K2 IV-V                & [5] \\ 
   $V$ [mag]                               &8.3   & 8.4571                   & [2] \\ 
   $B-V$ [mag]                             &0.945 & 0.981 $\pm0.013$         & [4] \\
   Mass    [$M_{\odot}$]                   &$0.70^{+0.04}_{-0.04}$ & $0.836^{+0.023}_{-0.025}$& [1] \\  
   Radius    [$R_{\odot}$]                  & - & $0.837^{+0.026}_{-0.018}$& [1] \\  
   Age    $[$Gyr$]$                        & -  & $10.71^{+2.24}_{-1.25}$  & [1] \\
   Distance  [pc]                          & - & $28.462^{+0.019}_{-0.016}$& [6] \\
   $\pi$  [mas]                            & 34.6 & $35.0393^{+0.0484}_{-0.0484}$      & [3] \\
   Luminosity    [$L{_\odot}$]              & 0.41 & $0.407^{+0.038}_{-0.031}$& [1] \\
   $T_{\mathrm{eff}}$~[K]                   & $4995^{+50}_{-50}$ & $5033.8^{+85.7}_{-77.2}$ & [1] \\
   $\log g~[\mathrm{cm\cdot s}^{-2}]$       & - & 4.23  $\pm0.07$          & [1] \\   
   {}[Fe/H]                                 & 0.15  $\pm0.06$ & 0.13  $\pm0.07$          & [1] \\ 
   
\hline \noalign{\vskip 0.5mm}   
\end{tabular}
\tablefoot{
1 -- This work.  2 -- \citet{Leeuwen}. 3 -- \citet{Gaia_Collaboration2018b}. 4 -- \citet{DaSilva2006}. 5 -- \citet{R.O.Gray2006}. 6 -- \citet{Bailer-Jones2021}.
}
\end{table}

\section{Stellar parameters}
\label{Sec3}

GJ\,1137 (HD\,93083, HIP\,52521) is a K2 IV-V dwarf star \citep{R.O.Gray2006} with an apparent magnitude of V=$8.46$ mag. The target was observed by HARPS and thus included in the {\sc HARPS-RVB}ank, which provides activity indicators and basic stellar atmospheric parameters. These parameters are estimated using templates created from the co-adding of HARPS spectra (see \citealp{Perdelwitz2024}, Sect.~2.1). Using these as a basis, we refined the stellar parameters of GJ\,1137 by analysing its spectral energy distribution (SED) with astroARIADNE\footnote{\url{https://github.com/jvines/astroARIADNE}} \citep{Vines&Jenkins2022}. This package employs a Bayesian approach, utilising various SED models to mitigate model-specific biases. It also estimates stellar mass and age by interpolating MESA isochrones and Stellar Tracks \citep[MIST,][]{Dotter2016, Choi2016}. As input, we used photometric measurements in the following bands: GALEX NUV, Johnson U, B, and V, Tycho B and V, Gaia DR2 G, BP, and RP, TESS, 2MASS J, H, and K, Wise W1 and W2. These were fitted with atmosphere models from PHOENIX v2 \citep{Husser2013}, BT-Settl, BT-NextGegn, BT-Cond \citep{Hauschildt1999, Allard2012}, Castelli \& Kurucz \citep{Castelli&Kurucz2003}, and Kurucz \citep{Kurucz1993}. 
We assign a normally distributed prior to the distance, using the estimate from Gaia EDR3 \citep{Bailer-Jones2021}. For interstellar extinction, we employ a flat prior ranging from zero to the maximum line-of-sight extinction. This was calculated using the galactic dust map by \citet{Schlafly&Finkbeiner2011} in the Python package dustmaps\footnote{\url{https://github.com/gregreen/dustmaps}} \citep{Green2018}.
For the initial stellar parameters T$_{\text{eff}}$, $\log{g}$, and [Fe/H], we constructed normal priors using the estimates of \citet{Perdelwitz2024} in the {\sc HARPS-RVB}ank. 

Our analysis demonstrates that GJ\,1137 has an age of $10.71^{+2.24}_{-1.25}$  Gyr, a mass of $0.836^{+0.023}_{-0.025}\,\unit\solarmass$ and a radius of $0.837^{+0.026}_{-0.018}\,\unit\solarradius$. The derived parameters are consistent with the latest data in the literature \citep[TICV8,][]{Stassun2017}. However, a significant variance is evident when comparing the stellar mass with the value used by \citet{Lovis2005}. The revised higher stellar mass implies an increased minimum mass for the known planet GJ\,1137 b, further discussed in \autoref{sec4.6}. A summary of all stellar parameters is presented in \autoref{table:phys_param}, and the resulting SED fit can be seen in \autoref{fig:SED}.

\begin{figure*}
    \centering
    \resizebox{18cm}{!}{\includegraphics{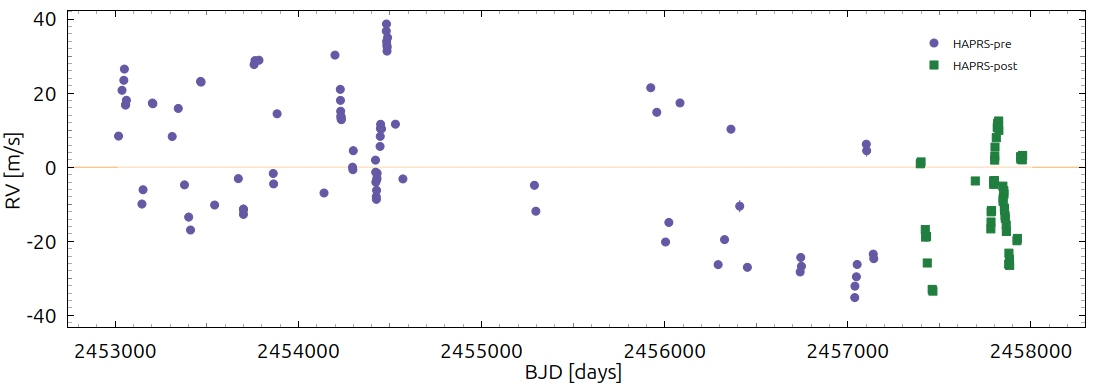}}
    \caption{
   HARPS RV time series for GJ\,1137. The data are divided into two subsets due to known instrumental RV offsets introduced by the HARPS optical fibre upgrade in May 2015. Purple circles represent RVs obtained before the fibre change, while green squares correspond to data taken after the upgrade. The observed RV variations are primarily driven by the signal of the known Saturn-mass exoplanet GJ\,1137\,b, with a period of $P = 144.7$\,d. 
   In addition, the RV data reveals a second, long-term signal from a potential Jovian planet with a minimum mass of approximately $1.3\,\unit\jupitermass$ and a period $P = 5640\pm240$\,d. Subsequent analysis attributes this long-term signal to stellar magnetic activity.
    }
    \label{fig:RVExo}
\end{figure*}
 
\section{Analysis and results}
\label{Sec4}

\subsection{Preliminary RV and stellar activity analysis}

Our preliminary data analysis relies on the {\tt Exo-Striker} exoplanet toolbox\footnote{\url{https://github.com/3fon3fonov/exostriker}} \citep{Trifonov2019_es}, which provides convenient exoplanet data analysis tools, but also direct access to {\sc HARPS-RVB}ank data. 
\autoref{fig:RVExo} shows the {\sc HARPS-RVB}ank SERVAL nightly zero-point corrected RV time series, which clearly exhibit at least two periodic signals:
a shorter-period signal with $P=144.75\pm0.028$ d attributed to the known exoplanet GJ\,1137\,b, and an additional long-period significant signal with $P = 5640\pm240$\,d , which sparked our interest in this system.

We performed a detailed time-series spectroscopic analysis to find that several critical activity indicators, including BIS, contrast, FWHM, dLW, and $\log R'_{\rm HK}$, exhibit significant low-frequency periodicities that closely resemble those found in the RV residuals after modelling the GJ\,1137\,b signal. In our local example, the Solar System, both Jupiter and the Sun's 11-year magnetic activity cycle induce RV variations on comparable timescales if the Sun were observed as a distant star. However, the disentanglement of these phenomena in RV is challenging, which raises caution regarding a possible stellar activity origin for the long-period variation.

As a next step in interpreting the long-term RV signal, we subtracted the contribution from the known planet and examined potential linear correlations between the residual RV variations and stellar activity indicators, using Pearson's correlation coefficient. We identified positive or negative correlations that exceed 0.25 in all available activity proxies, with the strongest correlation observed with the $R'_\text{HK}$ index (r = 0.49). This moderate positive correlation suggests that the long-term signal may be at least partially driven by magnetic activity rather than an additional planetary companion. To investigate this further, with greater statistical rigour, we conducted a more sophisticated analysis incorporating long-term functions (LTFs) and Gaussian processes (GPs) modelling.  Additional details and graphs on our preliminary analysis are provided in the Appendix.

\subsection{Combined orbital modelling including stellar activity}
\label{Sec4.2}

We use the modelling framework described in \citet{Stefanov2025a}, hereinafter \citetalias{Stefanov2025a}, which is generalised for $N$ datasets (e.g. HARPS-pre, HARPS-post; \mbox{$i\in[0..N-1]$}) and $M$ physical quantities (e.g. RV, FWHM; \mbox{$j\in[0..M-1]$}). The \citetalias{Stefanov2025a} framework includes:
(i) LTFs data that follow physical or instrumental trends;
(ii) offsets between different datasets;
(iii) dataset-dependent jitters for each physical quantity;
(iv) planetary RV fitting;
(v) stellar activity modelling using GPs. We now briefly recapitulate on GPs.

The activity originating from stellar rotation is regarded as a stochastic process, which must nevertheless be related to the stellar rotation period $P_\text{rot}$. A popular way to model this process is through GPs \citep{Rasmussen2006}, which assume a certain functional relationship of the correlation between measurements with respect to their difference in time. A common assumption for this correlation is that it is sensitive to $P_\text{rot}$, as well as some exponential decay in time, which can be naïvely ascribed to the active-region evolution on the stellar surface. This assumption gives rise to the squared-exponential periodic (SEP) kernel of the form
\begin{equation}
    k_\text{SEP}(\Delta t) = \kappa^2\exp\left[
    -\frac{\Delta t^2}{2\tau^2}
    -\frac{\sin^2\left(\pi\Delta t/P\right)}{2\eta^2}
    \right],
    \label{eq:sep_kernel}
\end{equation}
where $\kappa^2$ is the maximum amplitude of the kernel, $\tau$ is the timescale of coherence, $P$ is associated with the stellar-rotation period $P_\text{rot}$, and $\eta$ describes the kernel-feature complexity \citep{Haywood2014,Rajpaul2015,Angus2018}. We hereafter refer to $\eta$ as the `sinescale' because of its algebraic analogy to the timescale $\tau$.

Gaussian processes should be applied carefully, as they tend to overfit and absorb signals beyond $P_\text{rot}$ \citep{SuarezMascareno2023}, and the latter may not necessarily pertain to stellar rotation. One way to combat this overfitting is to constrain GPs by simultaneously fitting them on RVs with one or more activity indices \citep{Barragan2023}. For GJ\,1137, we found that fitting on the RV and FWHM time series alone gives a sufficiently well-described stellar-rotation activity. The size of our correlation matrix grows quadratically with the number of physical quantities, and the amount of additional information from the inclusion of $R'_\text{HK}$ as a second activity indicator does not justify the computational overhead. We restrict further GP models in this two-dimensional case; in these models, we shall follow the notation of \citetalias{Stefanov2025a} where \mbox{$j=0$} is assigned to RVs and \mbox{$j=1$} is assigned to the FWHM time series.

A common way to fit GPs on two-dimensional data follows the $FF'$ formalism that was defined in \citet{Aigrain2012} and extended in \citet{Rajpaul2015}. For measurements $y_{i,j}$ at time $t_{i,j}$, we solve the system of equations
\begin{equation}
    \left|\begin{array}{l}
    y_{i,0}=A_0\,G(t_{i,0})+B_0\,(\mathrm{d}G/\mathrm{d}t)_{t=t_{i,0}}\\[0.2ex]
    y_{i,1}=A_1\,G(t_{i,1})
    \end{array}\right.\qquad
    \begin{array}{l}
        \forall\, i,
    \end{array}
    \label{eq:multidimensional_system}
\end{equation}
where $A_j$ and $B_j$ are quantity-specific fit parameters, and $G(t_{i,j})$ is the kernel estimation of the stellar activity at given times. This regime is known to resolve degeneracies when there is a significant correlation between physical quantities \citep{SuarezMascareno2020}. We use the \textsc{s+leaf} MultiSeriesKernel procedure to realise it \citep{spleaf1,spleaf2}. For all models discussed in this work, we used \mbox{ReactiveNestedSampler}, a nested-sampling integrator provided by \textsc{UltraNest} \citep{ultranest}, in order to infer the posterior distribution of the parameters and numerically evaluate the Bayesian evidence $Z$ \citep{Skilling2004}. In every inference instance of $N_\text{param}$ model parameters, we required $40N_\text{param}$ live points and a region-respecting slice sampler that accepted $2N_\text{param}$ steps until the sample was considered independent.

\subsubsection{Joint analysis of RVs, FWHM and $R'_\text{HK}$}

\begin{figure*}
    \centering
    \resizebox{\hsize}{!}{\includegraphics{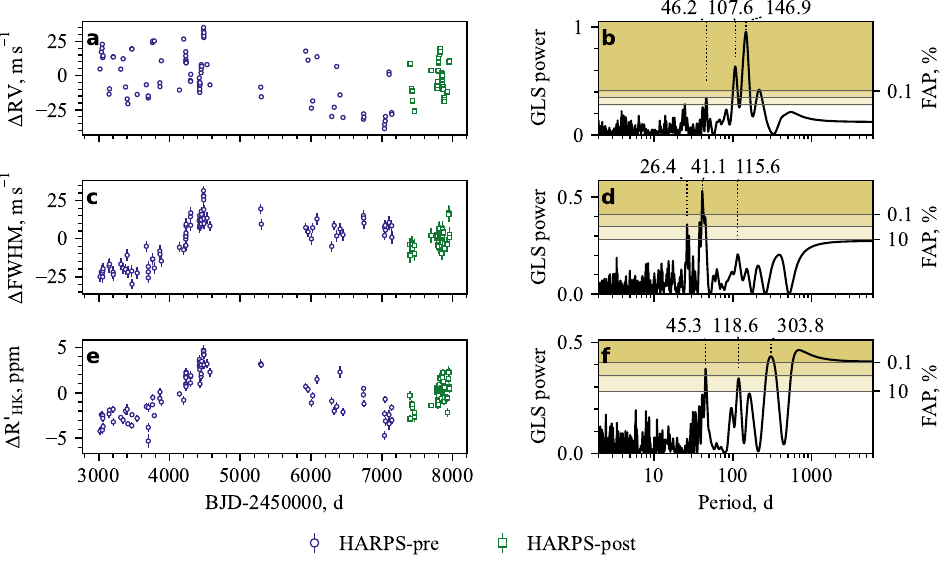}}
    \caption{
    Raw time series of mean-subtracted:
    (a) RV,
    (c) FWHM,
    (e) R$'_\text{HK}$. Measurements are marked depending on the data source: blue circles for HARPS-pre, and green squares for HARPS-post. (b,d,f) Associated wide-period GLSPs of RV and activity indicators. Three false alarm probability (FAP) levels: 10\%, 1\%, and 0.1\%, split GLSP ordinates in bands of different colour. We highlight the three most prominent peaks in each GLSP.
    }
    \label{fig:timeseries}
\end{figure*}

Figure \ref{fig:timeseries} shows the  RV, FWHM and $R'_\text{HK}$ raw-data time series and their generalised Lomb-Scargle periodograms (GLSPs; \citealp{Lomb1976, Scargle1982, Zechmeister2009, VanderPlas2015}) in the period range \SIrange{2}{6000}{\day}. Both the time series and GLSP panels offer a wealth of information about the behaviour of GJ\,1137. Firstly, the FWHM and the $R'_\text{HK}$ time series both show similar qualitative evolution, with a long-term sine-like component of a period near \SI{5000}{\day} (Figs.~\ref{fig:timeseries}a,c). In addition, both physical quantities appear to follow linear or higher-order polynomial trends. This is supported by their respective GLSPs (Figs.~\ref{fig:timeseries}d,f), which both show unresolved, yet significant, power excess at \SI{6000}{\day} and longer periods. We stress that a similar \SI{5000}{\day} cycle can be faintly discerned in the RV time series (Fig.~\ref{fig:timeseries}a), but these time series are dominated by GJ\,1137\,b 
(\SI{144}{\day}; \citealp{Lovis2005}), which is the main source of scatter, as well as the dominant signal in the GLSP (Fig.~\ref{fig:timeseries}b). We note the presence of additional peaks in all physical quantities (e.g. between \SIrange{41.1}{46.2}{\day}), but the long-term cycle must be modelled first to discuss those.

\subsubsection{Presence of a long-term stellar activity cycle}

We took all HARPS measurements in RV, FWHM, and $R'_\text{HK}$ and composed a three-dimensional model that contained a single elliptical-orbit planet and the following three-dimensional LTF:
\begin{equation}
    f_j(t_{i,j}) =
    \sum_{n=0}^{p}
    \alpha_{n,j} t^n_{i,j} +
    k_{\text{cyc, }j} \sin\left[\
    \frac{2\pi(t_{i,j}+P_\text{cyc}\,\varphi_{\text{cyc, }j})}{P_\text{cyc}}
    \right],
\end{equation}
where the first term models a $p$-dimensional polynomial trend with coefficients $\alpha_{n,j}$, and the second term is a common cycle of amplitude $k_{\text{cyc, }j}$ and phase $\varphi_{\text{cyc, }j}$, with a common cycle period $P_\text{cyc}$. That is to say, we restrict the long-term behaviour of each physical quantity to be modelled under a different polynomial, but we impose an additional sine term with a common period in all of them. No GPs were involved in this stage. We used the following period priors:
\mbox{$\mathcal{U}_{\log}\left(10^3, 10^4\right)\,\unit\day$} for the sine component in the LTF (i.e., the long-term cycle) and
\mbox{$\mathcal{U}\left(140, 150\right)\,\unit\day$} in accordance with the confirmed planetary companion GJ 1137 b.

HARPS-pre is known to have had a slowly drifting focus before the change of fibre. That manifested as a trend in the  FWHM data \citep{Zechmeister2013,Dumusque2018,Costes2021}. We found that a linear polynomial is enough to catch our FWHM trend in HARPS-pre \mbox{($i=0,~j=1$)}. Having accounted for this in the LTF as a separate linear-trend term $\beta$, our best solution in terms of $\Delta \ln Z$ is an LTF with a zero-order RV trend, a zero-order FWHM trend, and a quadratic $R'_\text{HK}$ trend. That is to say, our final choice of LTF is
\begin{equation}
    f_j(t_{i,j}) =
    \alpha_{0,j} +
    \delta_{i0}\delta_{j1}\beta t_{i,j}
    +
    k_{\text{cyc, }j} \sin\left[\
    \frac{2\pi(t_{i,j}+P_\text{cyc}\,\varphi_{\text{cyc, }j})}{P_\text{cyc}}
    \right],
    \label{eq:ltf_cycle}
\end{equation}
where $\delta$ is the Kronecker delta function. Table~\ref{tab:model_cycle_posterior} lists all parameters in the model, together with their adopted priors and inferred posteriors. We measure a cycle period of $5320^{+170}_{-150}\,\unit\day$ with a well-behaving unimodal posterior shape. We obtain the following RV, FWHM, and $R'_\text{HK}$ semi-amplitudes of the cycle:
$14.2\pm 0.6\,\unit{\metre\per\second}$,
$18.7^{+1.8}_{-1.7}\,\unit{\metre\per\second}$, and
$6.59^{+0.89}_{-0.79}\,\unit\ppm$ respectively -- all of them standing at least $8\sigma$ away from the zero. These results underline a strong \SI{15}{\year} cycle in RV and activity. The cycle phase somewhat differs for different quantities -- we obtain RV, FWHM, and $R'_\text{HK}$ phases of
$0.894^{+0.020}_{-0.020}$,
$0.888^{+0.021}_{-0.022}$, and
$0.797^{+0.021}_{-0.020}$ respectively. RV and FWHM are in phase within $1\sigma$. The cycle appears to lag slightly behind in $R'_\text{HK}$. While such phase lags are observed for the Sun \citep{CollierCameron2019} and other stars \citep{Burrows2024}, our posterior may be deceiving, since this baseline model does not account for other sources of stellar variability (e.g. stellar rotation with GPs).

\begin{figure*}
    \centering
    \resizebox{\hsize}{!}{\includegraphics{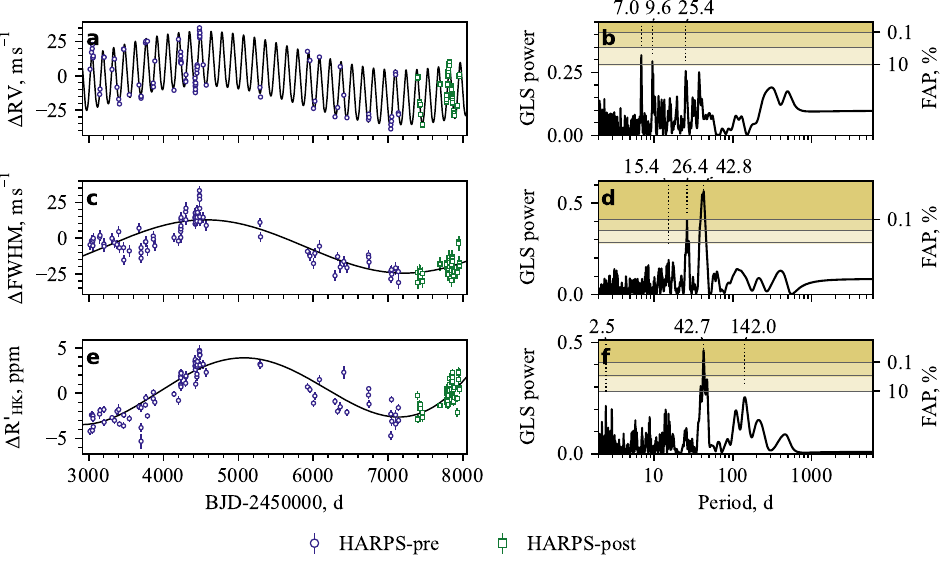}}
    \caption{
    Time series (markers) against our LTF model (black), over the whole dataset baseline in:
    (a) RV,
    (c) FWHM,
    (e) R$'_\text{HK}$. We implicitly correct for the HARPS-pre FWHM drift before visualising. (b,d,f) Associated GLSPs of the residual time series, accounting for the model jitter .
    }
    \label{fig:timeseries_cycle}
\end{figure*}

Figure~\ref{fig:timeseries_cycle} shows the model fit against data, as well as the periodograms of the residuals, which accounted for the model jitter. The modelling of the long-term cycle and GJ\,1137\,b was enough to cause all significant peaks beyond \SI{100}{\day} to disappear. On the short-period end, we see that a \SI{43}{\day} signal remained significant in FWHM and $R'_\text{HK}$. This would be consistent with our expectations for the rotational period $P_\text{rot}$ -- the mean and standard deviation of our $\log_{10}R'_\text{HK}$ measurements form an estimate of $-4.98\pm 0.10$, which translates as
\mbox{$P_\text{rot}=31\pm 7\,\unit\day$}
through the relation in \citet{SuarezMascareno2015}. The FWHM GLSP displays a strong \SI{26.4}{\day} signal that may be the first harmonic of $P_\text{rot}$. With regard to RV, we notice two peaks in \SI{7.0}{\day} and \SI{9.6}{\day} that soared to <10\% FAP (\citealp{Baluev2008}). 

\subsubsection{Presence of short-term stellar activity and an additional RV signal}

We continued with the aforementioned LTF model and augmented it with a stellar-activity component as described in Sect.~\ref{Sec4.2}. We chose to test for four approximations of the SEP kernel that \textsc{s+leaf} offers: the Matérn 3/2 exponential periodic (MEP) kernel and the exponential-sine periodic (ESP) kernel with two, three or four harmonics (hereafter ESP2, ESP3, ESP4). We additionally tested for a second planetary signal with a period prior of
\mbox{$\mathcal{U}_{\log}\left(1, 100\right)\,\unit\day$}
and a semi-amplitude prior of
\mbox{$\mathcal{U}\left(0,10\right)\,\unit{\metre\per\second}$}. For this second potential RV contribution, we independently tested for a circular and a Keplerian orbit. The variation of four stellar-activity kernels (MEP, ESP2, ESP3, ESP4) and the three planetary configurations (Keplerian, Keplerian+circular, Keplerian+Keplerian) resulted in a grid of 12 models. These models were fitted on the conjunction of RV and FWHM data. We modelled stellar activity in the multi-dimensional regime, with RV having a gradient with respect to FWHM. We used our aforementioned rotation-period estimate through \citet{SuarezMascareno2015} \mbox{($31\pm 7\,\unit\day$)} to impose a rotation-period prior
\mbox{$P_\text{rot}=\mathcal{U}\left(10, 52\right)\,\unit\day$}
and the uninformed timescale prior
\mbox{$\tau=\mathcal{U}_{\log}\left(10, 10^4\right)\,\unit\day$}.

\begin{figure}
    \centering
    \resizebox{\hsize}{!}{\includegraphics{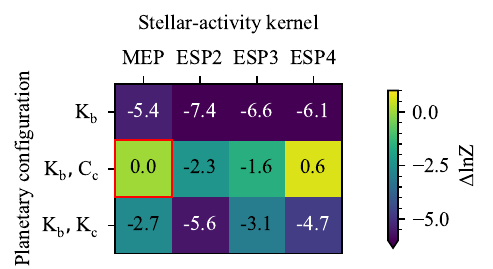}}
    \caption{
    Bayesian evidence comparison between different planetary configurations and stellar-activity kernels. Planetary configurations include circular orbit (C) and Keplerian (K) components. For stellar activity, we utilised the MEP kernel and the ESP kernel with 2, 3, and 4 harmonics. The model that we further elected for analysis assumed a model with a Keplerian signal, a circular signal, and an MEP kernel (red border; $\ln Z=-807.3$).
    We give the Bayesian factor $\Delta\ln Z$ of remaining models relative to this model.
    }
    \label{fig:model_table}
\end{figure}

Figure~\ref{fig:model_table} provides a comparison of models in our grid search using Bayesian evidence $\ln Z$. Entries are arranged so that planetary configurations increase in complexity downwards, while stellar-activity kernels increase in complexity rightwards. In the figure, we highlight the most appropriate model: an MEP kernel, with a Keplerian component for GJ\,1137\,b and an additional circular-orbit signal. What follows is the justification for this selection.

The average improvement across kernels from a one-Keplerian model is:
\mbox{$\Delta\ln Z=5.6$} for an additional circular-orbit signal and
\mbox{$\Delta\ln Z=2.4$} for an additional Keplerian signal. In terms of specific kernels, we get the following $\Delta\ln Z$ improvements for an additional circular-orbit signal:
5.4 for MEP,
5.1 for ESP2,
5.0 for ESP3,
6.7 for ESP4.
All of these improvements are great enough to be interpreted as overwhelming evidence of a second RV signal in the data (\mbox{$\ln Z>5$}; \citealt{Jeffreys1946}). We compare the period, phase, semi-amplitude, eccentricity, and argument-of-periastron posteriors of the second signal in Tables
\ref{tab:model_comparison_planetc_period},
\ref{tab:model_comparison_planetc_phase},
\ref{tab:model_comparison_planetc_krv},
\ref{tab:model_comparison_planetc_eccentricity}, and
\ref{tab:model_comparison_planetc_omega} respectively. These tables demonstrate high posterior stability across various planetary configurations and stellar-activity kernels. All posteriors agree within $1\sigma$ for any configuration-kernel pair. In two-Keplerian models, none of the inferred eccentricity posteriors is well defined with respect to zero (Table~\ref{tab:model_comparison_planetc_eccentricity}). This is consistent with the slightly lower evidence relative to Keplerian+circular models in Fig.~\ref{fig:model_table}, and reinforces that the new RV signal is indeed sine-like.

We checked if GP parameters were consistent across the model grid as well. Figure~\ref{fig:kernel_comparison} visually compares the posteriors of all GP parameters and the average RV jitter per point between models. The posteriors are qualitatively the same across planetary configurations and kernels, with the following exceptions. The rotation periods change median from about \SI{30}{\day} to about \SI{32}{\day} with the inclusion of a second RV signal. This is not too surprising, given that the period of the latter is of the same order (near \SI{9.6}{\day}). The ESP4 models in particular tend to struggle with bimodality, as they try to catch a weak $P_\text{rot}$ mode near \SI{40}{\day} (Figs.~\ref{fig:kernel_comparison}a,o). However, MEP-kernel models tend to overestimate timescales (Figs.~\ref{fig:kernel_comparison}b,i,p). At the same time, they infer a smaller RV jitter than the ESP family (Figs.~\ref{fig:kernel_comparison}g,n,u).

We assessed the significance of the \SI{9.6}{\day} RV signal following \citet{Hara2022}. Their formalism uses the likelihood of vectors drawn from the posterior distribution in order to compute the probability that there are no planets within a given orbital-frequency element. The authors refer to this metric as the false inclusion probability (FIP). We re-ran a derivative of our selected model with three planet components: the same Keplerian that captures GJ\,1137\,b, as well as two circular-orbit components that share the period prior \mbox{$\mathcal{U}_{\log}(1,100)\,\unit\day$}. We computed the FIP between \SIrange{1}{100}{\day} for an angular-frequency step \mbox{$\Delta\omega=2\pi/5T$}, where $T$ is the total time series baseline. The novel \SI{9.6}{\day} RV signal is assigned a FIP smaller than \num{e-5}, which is much more significant than the \num{e-2} threshold proposed by \citet{Hara2022}. We found no other RV peaks with significant FIPs.

Then, we tested whether the \SI{9.6}{\day} RV signal is not simply a power excess that our GP kernels could not absorb. We followed the treatment of \citetalias{Stefanov2025a}, Sect. 4.5.3. We started with our selected model (Keplerian+circular MEP), removed the circular-orbit RV component, and instead augmented an extra sine term in the LTF (Equation~\ref{eq:ltf_cycle}) that acts on both the RV and the FWHM. In this way, we can still model for a long-term cycle with the original sine term in the LTF -- and at the same time try to guide the second sine term \SI{9.6}{\day} to test whether there is a significant signal in the FWHM too. This derivative model used the same wide period prior $\mathcal{U}_{\log}(1,100)\,\unit\day$. Its evidence scores worse by
\mbox{$\Delta\ln Z=-2.3$} relative to our selected model. The short-term cycle designed to catch the \SI{9.6}{\day} RV signal actually locates it (Fig.~\ref{fig:pseudoplanet_posterior}). The \SI{9.6}{\day} signal is assigned a FWHM semi-amplitude of \mbox{$0.62^{+0.66}_{-0.44}\,\unit{\metre\per\second}$}, a measurement that is statistically consistent with zero and that is significantly lower than the median FWHM uncertainty of the dataset (\SI{4.01}{\metre\per\second}). On that account, we conclude that the \SI{9.6}{\day} signal manifests in RV only and is not associated with stellar activity.

Finally, we verified the stability of the \SI{9.6}{\day} RV signal through an apodisation test \citep{Gregory2016, Hara2022a}. We re-ran the best model (Keplerian+circular MEP), but with the following modification in the semi-amplitude of the circular component:
\begin{equation}
    k_\text{rv, c} \to k_\text{rv, c}
    \exp\left[-\frac{(t'-\mu_\text{apo})^2}{2\sigma_\text{apo}^2}\right],
\end{equation}
that is,  made it follow a Gaussian-like evolution in time. A true planetary signal is stable over time, meaning that a poorly-constrained $\mu_\text{apo}$ and a large $\sigma_\text{apo}$ would pass this planetary-nature test. We used a uniform prior for $\mu_\text{apo}$ that covers the HARPS baseline, the uninformed
\mbox{$\sigma_\text{apo}=\mathcal{U}_\text{log}\left(1,10^6\right)\,\unit\day$};
lastly, we constrained the orbital period prior to
\mbox{$P_\text{c}=\mathcal{U}\left(9,10\right)\,\unit\day$} to help inference.
The $\mu_\text{apo}$ posterior was indistinguishable from a uniform distribution. The $\sigma_\text{apo}$ showed a similar characteristic in log-space, with a step function that had a transition near the HARPS baseline (near \SI{4900}{\day}). We report
\mbox{$\sigma_\text{apo}/T_\text{baseline}=10.6^{+69.6}_{-9.2}$}.

Figure~\ref{fig:apo_signal} shows the posterior of the apodised semi-amplitude $k_\text{rv, c}$ over time, in $1\sigma$, $2\sigma$ and $3\sigma$ confidence intervals. We observe an amplitude that is consistent with the posterior of our best-model
\mbox{($1.73^{+0.24}_{-0.23}\,\unit{\metre\per\second}$; Table~\ref{tab:bestmodel_posterior})}.
The apodised $k_\text{rv, c}$ begins to demonstrate variance only in the $3\sigma$ confidence interval. The upper $3\sigma$ bound reaches \SI{3}{\metre\per\second} in the middle of the baseline. We attribute this to the scarcity of observations in the middle of the baseline, which leaves some degree of freedom to the Gaussian profile.

\subsection{The GJ\,1137 multi-planetary system}
\label{sec4.6}

\begin{figure*}
    \centering
    \resizebox{\hsize}{!}{\includegraphics{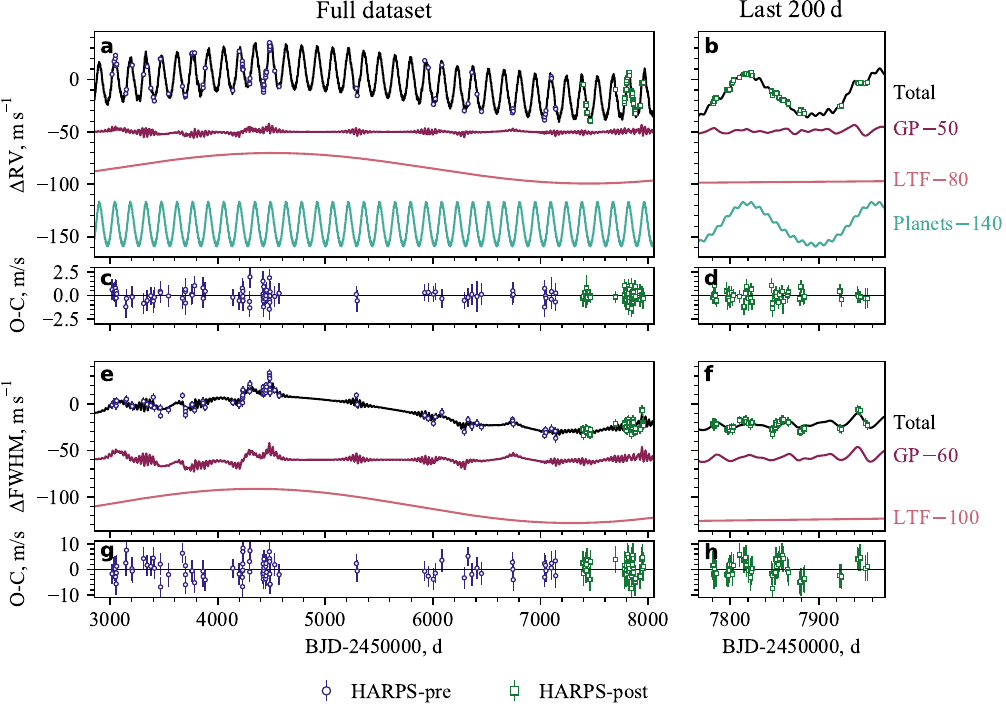}}
    \caption{
    Time series (error bars) against our best model (solid lines) over the full dataset and over a selected temporal region. Data points come with error bars that account for the instrument and the model jitter.
    (a,b) SERVAL RV, where the full model (black) is split into the stellar-activity GP component (red), the LTF (pink), and the two-planet component (teal).
    (c,d) Associated RV residuals.
    (e,f) RACCOON FWHM, where the full model (black) is split into the stellar-activity GP component (red) and the LTF (pink).
    We implicitly correct for the HARPS-pre FWHM drift before visualising. Every component is offset by an arbitrary amount, labelled in the plots.
    (g,h) Associated FWHM residuals.
    }
    \label{fig:bestmodel_fit}
\end{figure*}

Figure~\ref{fig:bestmodel_fit} shows the results of our selected model. Before us stands a system that exhibits complex long-term and short-term behaviour, with two planetary signals additionally permeating the RV time series: those are GJ\,1137\,b \citep{Lovis2005}, and our new discovery, GJ\,1137\,c. Figure~\ref{fig:bestmodel_fit} highlights a good agreement between our selected model and data, both for the entire baseline and in a dense selected interval containing approximately the last \SI{200}{\day} of measurements. The temporal sampling of measurements is non-trivial, but the model appears to handle it effectively, without overfitting the data. This is supported by the spread of RV and FWHM residuals -- we observe no apparent trends with time, and their GLSPs show no other significant peaks (Fig.~\ref{fig:bestmodel_pgram}).
We report the following RV improvement in root mean square from raw time series to model residuals:
\SI{18.71}{\metre\per\second} to \SI{0.62}{\metre\per\second} in HARPS-pre, and
\SI{12.26}{\metre\per\second} to \SI{0.51}{\metre\per\second} in HARPS-post. When it comes to FWHM, these figures are
\SI{15.33}{\metre\per\second} to \SI{3.36}{\metre\per\second} in HARPS-pre, and
\SI{6.03}{\metre\per\second} to \SI{2.98}{\metre\per\second} in HARPS-post.

\begin{figure*}
    \centering
    \resizebox{\hsize}{!}{\includegraphics{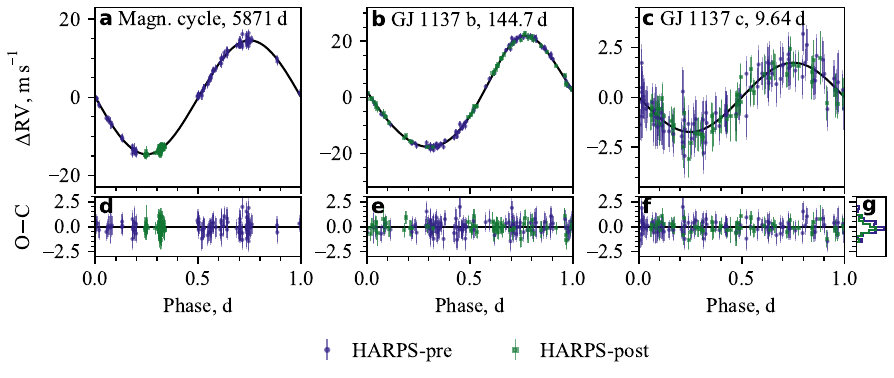}}
    \caption{
    (a,b,c) Activity-corrected, phase-folded time series (error bars) against the LTF- and the planetary fit of our best model. (d,e,f) Residual time series after the fit. (g) Distribution of residuals (solid lines).
    }
    \label{fig:bestmodel_children_signal}
\end{figure*}

Figure~\ref{fig:bestmodel_children_signal} shows the phase-folded plot of the signals of GJ\,1137\,b and GJ\,1137\,c after subtracting the activity-related components. The RV signal of the new planet appears to be well supported by the phased data.
Table~\ref{tab:bestmodel_posterior} displays the posteriors of our selected model. Having accounted for stellar activity and the two planets, we obtain a cycle period of $5870^{+480}_{-350}\,\unit\day$. We obtain the following RV and FWHM semi-amplitudes of the cycle:
\mbox{$14.6^{+1.3}_{-1.2}\,\unit{\metre\per\second}$}, and
\mbox{$18.4^{+3.5}_{-3.9}\,\unit{\metre\per\second}$} respectively. These results again support a strong \SI{15}{\year} activity in RV and activity. We obtain nearly matching cycle phases between dimensions:
\mbox{$0.837^{+0.037}_{-0.043}$} for RV and
\mbox{$0.864^{+0.054}_{-0.070}$} for FWHM. Our dataset assigns the following jitters to the HARPS-pre and HARPS-post datasets:
\mbox{$0.63^{+0.38}_{-0.37}\,\unit{\metre\per\second}$} and
\mbox{$0.29^{+0.26}_{-0.20}\,\unit{\metre\per\second}$} for RV, as well as
\mbox{$1.97^{+1.09}_{-1.14}\,\unit{\metre\per\second}$} and
\mbox{$0.67^{+0.75}_{-0.48}\,\unit{\metre\per\second}$} for FWHM.

In terms of stellar activity, we successfully constrain the stellar rotation to
\mbox{$P_\text{rot}=32.3^{+1.2}_{-1.3}\,\unit\day$} and derive a timescale of coherence
\mbox{$\tau=73^{+33}_{-21}\,\unit\day$}.
We obtain a sine scale of $0.85^{+0.26}_{-0.17}$ and the following GP amplitudes:
\mbox{$A_0=2.20^{+0.53}_{-0.42}\,\unit{\metre\per\second}$},
\mbox{$B_0=19.9^{+6.9}_{-4.7}\,\unit{\metre\per\second}$}, and
\mbox{$A_1=6.99^{+1.33}_{-0.99}\,\unit{\metre\per\second}$}.
All of them are positive, which suggests that stellar-activity modulations correlate positively between RV and FWHM. Furthermore, $A_0\ll B_0$, hinting that in GJ\,1137, stellar activity manifests more prominently in the RV gradient, which in turn is likely caused by a flux effect. The positive value of $B_0$ indicates that the active regions that we model are probably dark spots on the stellar surface.

With respect to GJ\,1137\,b, we significantly improve the orbital solution first given by \citet{Lovis2005}. We obtain a period of
\mbox{$P_\text{b}=144.720^{+0.029}_{-0.029}\,\unit\day$}, a semi-amplitude of
\mbox{$k_\text{rv, b}=19.8\pm 0.4\,\unit{\metre\per\second}$}, and an eccentricity of $e_b=0.118^{+0.016}_{-0.015}$. Our measurement for $k_\text{rv, b}$ differs by more than $3\sigma$ from \citet{Lovis2005}
\mbox{$(18.3\pm 0.5\,\unit{\metre\per\second})$}. From our model parameters, we derived that GJ\,1137\,b has a semi-major axis of
\mbox{$a_\text{b}=0.508\pm 0.05\,\unit\au$}, a minimum mass of
\mbox{$m_\text{b}\sin i_\text{b}=0.451\pm 0.012\,\unit\jupitermass$} (against $0.37\,\unit\jupitermass$ in \citealp{Lovis2005}), and a temporal-average instellation flux of
\mbox{$\Phi_\text{b}=1.59\pm 0.15\,\unit\earthflux$} as defined in \citet{Mendez2017}. Our new detection, GJ\,1137\,c, has a period of
\mbox{$P_\text{c}=9.6412^{+(12)}_{-(11)}\,\unit\day$} and a semi-amplitude of
\mbox{$k_\text{rv, c}=1.73^{+0.24}_{-0.23}\,\unit{\metre\per\second}$}, standing more than $6\sigma$ away from zero. For this planet, we derive a semi-major axis of
\mbox{$a_\text{c}=0.0835^{+(8)}_{-(8)}\,\unit\au$}, a minimum mass of
\mbox{$m_\text{c}\sin i_\text{c}=5.12^{+0.70}_{-0.69}\,\unit\earthmass$}, and an instellation flux of
\mbox{$\Phi_\text{b}=58.4^{+5.6}_{-5.5}\,\unit\earthflux$}.

\begin{figure}
    \centering
    \resizebox{\hsize}{!}{\includegraphics{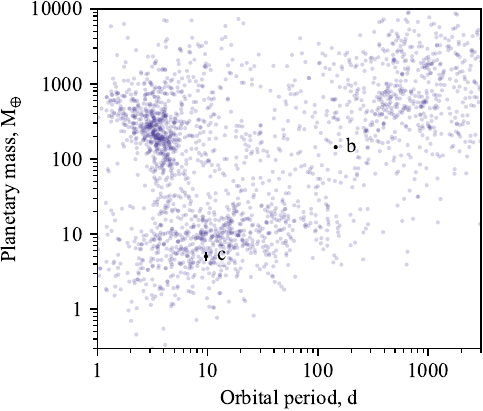}}
    \caption{
    Mass-period diagram of exoplanets with well-established masses and periods from \citet{nasaexoplanettable1,nasaexoplanettable2}. Against those, we plot the derived posteriors of our best model. Reported uncertainties reflect the 16\textsuperscript{th} and the 84\textsuperscript{th} percentiles.}
    
    \label{fig:bestmodel_children_msini}
\end{figure}

\begin{figure}
    \centering
    \resizebox{\hsize}{!}{\includegraphics{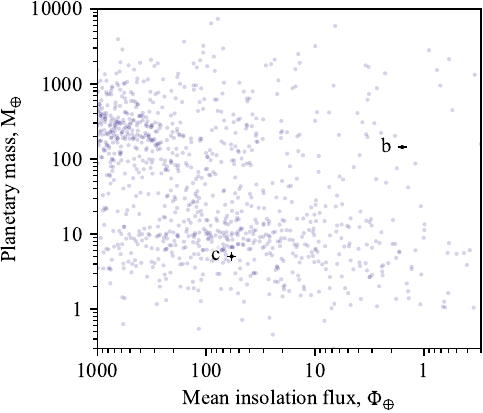}}
    \caption{
    Mass-flux diagram of exoplanets with well-established masses and periods from \citet{nasaexoplanettable1,nasaexoplanettable2}. Against those, we plot the derived posteriors of our best model. Reported uncertainties reflect the 16\textsuperscript{th} and the 84\textsuperscript{th} percentiles.}
    
    \label{fig:bestmodel_children_flux}
\end{figure}

\begin{figure}
    \centering
    \resizebox{\hsize}{!}{\includegraphics{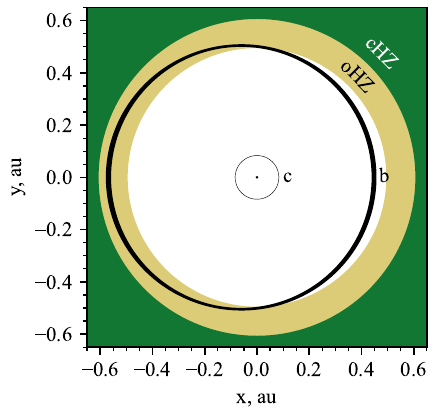}}
    \caption{
    Planetary system of GJ\,1137 relative to the conservative habitable zone (cHZ; green) and the optimistic habitable zone (oHZ; yellow), respectively. Black orbital bands come from our $1\sigma$ posteriors of the orbital parameters of GJ\,1137\,b and GJ\,1137\,c.
    }
    \label{fig:bestmodel_orbits}
\end{figure}

We computed the mass-period and the mass-flux relationships of known exoplanets that have had their minimum masses measured independently and that have relative mass uncertainties smaller than one third \citep{nasaexoplanettable1,nasaexoplanettable2}.
Figures~\ref{fig:bestmodel_children_msini} and \ref{fig:bestmodel_children_flux} display these diagrams for
the period range \SIrange{1}{3000}{\day},
the flux range \SIrange{0.3}{10000}{\earthflux}, and
the mass range \SIrange{0.3}{10000}{\earthmass} against our posteriors of the two planets in the GJ\,1137 system. GJ\,1137\,c lies in the bulk of the distribution of hot sub-Neptunes, while GJ\,1137\,b stands in the less explored regions of the parameter space in both figures. GJ\,1137\,b is particularly interesting because it receives an Earth-like average insolation flux \mbox{($1.59\pm 0.15\,\unit\earthflux$)}. We attempted to assess the habitable zone (HZ) of GJ\,1137 through \citet{Kopparapu2014}. Their work provides the runaway and maximum-greenhouse flux limits, which we take to define a conservative HZ (cHZ); as well as recent-Venus and early-Mars flux limits, which we take to define an optimistic HZ (oHZ). We used the median values of $M_\star$, $L_\star$, $T_\text{eff}$ from Table~\ref{table:phys_param} and estimated that for \SI{5}{\earthmass} planets, the cHZ extends in the range \SIrange{0.61}{1.14}{\au}, while the oHZ extends around it in the range \SIrange{0.50}{1.21}{\au}. Figure~\ref{fig:bestmodel_orbits} compares this HZ with the orbital bands of GJ\,1137\,b and GJ\,1137\,c that come from the posteriors of their orbital parameters.

According to commonly accepted theories of planet formation, the Saturn-mass planet GJ\,1137\,b most likely formed beyond the ice line of its host star and migrated inward via convergent migration \citep[][]{Lin1986,Beauge2003}. Consequently, GJ\,1137\,b may host a primordial system of icy exomoons, analogous to those around the giant planets in the Solar System. Such exomoons could have settled in stable warm orbits and may be potentially habitable, ocean-like, or early-Mars-like worlds \citep[see, discussion in e.g.][]{Trifonov2020,Hobson2023}. Although exomoon detection remains challenging with current state-of-the-art techniques, GJ\,1137\,b is an intriguing target for future analysis.

\subsection{Detection limits}
We assessed our detection limits in the following manner. We took the period interval \mbox{$\mathcal{U}_{\log}(1,1000)\,\unit\day$} and split it into 100 bins. In each of these bins, we took our best model and included an extra third circular-orbit planet with an $\mathcal{U}_{\log}$ orbital-period prior defined by that bin and an RV semi-amplitude of \mbox{$\mathcal{U}(0,5)\,\unit{\metre\per\second}$}. We fixed all other parameters except the RV zero-order correction $\alpha_{0,0}$ and the RV jitters $J_{i,0}$. Then, for each bin, we ran an independent simpler `bin model', inferred the posteriors of free parameters, and considered the $\pm 3\sigma$ percentiles of the RV semi-amplitude posterior of the sampled extra planet. To speed up inference, in all  instances of $N_\text{param}$ bin-model parameters, we required $20N_\text{param}$ live points (instead of $40N_\text{param}$ on the model grid).

\begin{figure}
    \centering
    \resizebox{\hsize}{!}{\includegraphics{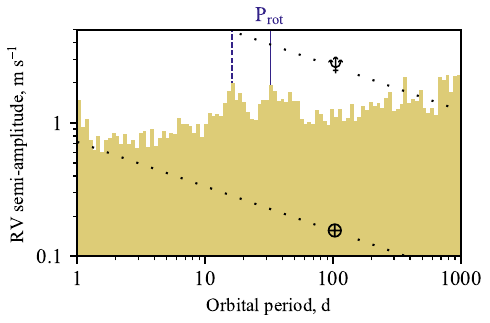}}
    \caption{
    Detection limits of our study. Shaded bands show the $+3\sigma$ RV semi-amplitude posterior of a modelled planet of a third potential planet added to the best model, in 100 log-spaced orbital-period bins. Stellar rotation somewhat degrades detection for orbital periods near $P_\text{rot}$ (solid blue line) and $P_\text{rot}/2$ (dashed blue line). Dotted lines show the RV semi-amplitudes that an Earth-mass planet and a Neptune-mass planet would inject in the system RV.
    }
    \label{fig:bestmodel_detectability}
\end{figure}

Figure~\ref{fig:bestmodel_detectability} displays the $+3\sigma$ percentiles of our 100 bins in the parameter space.
Overall, our $+3\sigma$ detection limit gradually increases from \SI{1}{\metre\per\second} to \SI{2}{\metre\per\second} over  orbital-period space. There are two apparent degradations of \SI{2}{\metre\per\second} in orbital periods close to $P_\text{rot}$ and $P_\text{rot}/2$, which is expected from the limitations of GP frameworks. We are sensitive to the detection of any Neptune-mass planets up to about \SI{300}{\day}. Typically, $-3\sigma$ percentiles are indicative of subtle yet significant signals that merit attention in subsequent studies. In our case, we do not observe $-3\sigma$ percentiles above \SI{10}{\centi\metre\per\second}.

\section{Summary and conclusions}
\label{Sec5}

We report the results of a comprehensive analysis of precision radial velocity data and stellar activity indicators for the nearby K-dwarf star GJ\,1137 (HD\,93083, HIP\,52521). Our study is based on 140 archival HARPS spectra that span more than 13 years and provide a long and consistent temporal baseline to investigate the presence of additional Jovian-mass analogs in this system, beyond the already known Saturn-mass exoplanet GJ\,1137\,b discovered by \citet{Lovis2005}.

We refine the stellar parameters of GJ\,1137 and report a stellar mass of $0.836^{+0.023}_{-0.025}\,\unit\solarmass$. Incorporating this updated mass and the extended RV dataset collected since 2005, we revised the orbital parameters of the Saturn-mass planet GJ\,1137\,b. We determine a minimum mass of $0.451^{+0.012}_{-0.012}\,\unit\jupitermass$ and an orbital period of $P_b = 144.7$\,d, which places GJ\,1137\,b near the optimistic habitable zone of its host star. Due to its location and mass, GJ\,1137\,b may be of particular interest in the context of habitability, as it could host Galilean-like exomoons with stable orbits that receive moderate insolation, making them potentially habitable.

Our objective is to discover and characterise Jovian-mass planets orbiting low-mass stellar hosts. GJ\,1137 was of particular interest because, in addition to the known RV signal induced by GJ\,1137\,b, we identified a second long-period RV variation initially suggestive of a distant Jovian companion. However, a detailed analysis of spectroscopic activity indicators—including RV, FWHM, and $\log R'_{\rm HK}$, among other activity indices -- revealed a strong common period at $P_\text{cyc} = 5870^{+480}_{-350}\,\unit\day$. This led us to conclude that the observed RV signal is most likely induced by a stellar magnetic activity cycle with a period of $P_\text{cyc}$, akin to the 11-year solar activity cycle, rather than by a Jovian-mass planet. We therefore attribute the long-period RV variation to stellar activity.

After correcting for long-term stellar activity, we uncovered a short-period planetary candidate, GJ\,1137\,c, with an orbital period of $9.64$\,d and a minimum mass of $5.12^{+0.70}_{-0.69}$\,M$_\oplus$, placing it firmly in the super-Earth regime. This signal is consistently recovered across various stellar noise models and kernel configurations, and we find no significant correlation with any activity indicators. FIP tests, following \citet{Hara2022}, and an apodisation of the signal associated with GJ\,1137\,c, following \citet{Hara2022a}, further support the planetary nature of this signal.

Our findings illustrate the complexities and potential pitfalls in interpreting long-period RV trends, particularly in magnetically active stars. This is especially relevant for the detection of Jovian analogues, whose occurrence rates are estimated at $\sim$3--6\% around FGK stars \citep[e.g.][]{Wittenmyer2020,Rosenthal2021}, and are likely even lower for M and K dwarfs. In this context, GJ\,1137 serves as a cautionary example, where long-term activity signals and sparse sampling could mimic planetary companions.

Overall, GJ\,1137 emerges as a dynamically rich multi-planetary system. The presence of a close-in super-Earth, a longer-period Saturn-mass planet, and a prominent magnetic activity cycle makes it a compelling target for continued monitoring. Our results emphasise the importance of joint RV and activity diagnostics in high-precision Doppler surveys, particularly when searching for long-period, low-amplitude planetary signals.

\section*{Data availability}

\autoref{tab:RVs} is fully available in electronic form at the CDS via anonymous ftp to cdsarc.u-strasbg.fr (130.79.128.5) or via http://cdsweb.u-strasbg.fr/cgi-bin/qcat?J/A+A/.

\begin{acknowledgements}
We thank the anonymous referee for their constructive comments that helped to improve the quality of this work.
This research is based on spectroscopic observations made with ESO's 3.6m telescope at La Silla Observatory under programme IDs 072.C-0488, 183.C-0972, 091.C-0936, 192.C-0852, 198.C-0836. We gratefully acknowledge Michel Mayor, Stéphane Udry, and Rodrigo Díaz, principal investigators of the HARPS programme, for providing access to the observational data used in this work.
This research has made use of the SIMBAD database, operated at CDS, Strasbourg, France.
This work has made use of data from the European Space Agency (ESA)
mission {\it Gaia} (\url{https://www.cosmos.esa.int/gaia}), processed by
the {\it Gaia} Data Processing and Analysis Consortium (DPAC,
\url{https://www.cosmos.esa.int/web/gaia/dpac/consortium}). Funding
for the DPAC has been provided by national institutions, in particular
the institutions participating in the {\it Gaia} Multilateral Agreement.
D.S., S.S, E.Z., D.A., V.B., and T.T. acknowledge support by the BNSF program "VIHREN-2021" project No. KP-06-DV/5.
This research was in part funded by the UKRI Grant EP/X027562/1.
A.K.S. acknowledges the support of a fellowship from the ``la Caixa'' Foundation (ID 100010434). The fellowship code is LCF/BQ/DI23/11990071.
A.K.S., J.I.G.H., A.S.M., R.R., and N.N. acknowledge financial support from the Spanish Ministry of Science, Innovation and Universities (MICIU) project PID2023-149982NB-I00.
N.N. acknowledges funding from Light Bridges for the Doctoral Thesis ``Habitable Earth-like planets with ESPRESSO and NIRPS'', in cooperation with the Instituto de Astrofísica de Canarias, and the use of Indefeasible Computer Rights (ICR) being commissioned at the ASTRO POC project in Tenerife, Canary Islands, Spain. The ICR-ASTRONOMY used for his research was provided by Light Bridges in cooperation with Hewlett Packard Enterprise (HPE).

We used the following \textsc{Python} packages for data analysis and
visualisation:
\textsc{Exo-Striker} \citep{Trifonov2019_es},
\textsc{Matplotlib} \citep{matplotlib},
\textsc{nieva} (Stefanov et al., in prep.),
\textsc{NumPy} \citep{numpy},
\textsc{pandas} \citep{pandas1,pandas2},
\textsc{SciPy} \citep{scipy},
\textsc{s+leaf} \citep{spleaf1,spleaf2}, and
\textsc{ultranest} \citep{ultranest}.
The bulk of modelling and inference was done on
the Diva cluster (192 Xeon E7-4850 \SI{2.1}{\giga\hertz} CPUs; \SI{4.4}{\tera\byte} RAM) at Instituto de Astrofísica de Canarias, Tenerife, Spain.

\end{acknowledgements}

\bibliographystyle{aa} 
\bibliography{GJ1137_bibtex}

\let\cleardoublepage\clearpage
\begin{appendix} 
\label{appendix}

\section{Supplementary material}

\subsection{Preliminary spectroscopic analysis with the Exo-Striker and MLP}

We performed a periodogram analysis of the SERVAL radial velocities and activity index time series using the maximum likelihood periodogram \citep[MLP;][]{Baluev2008, Baluev2009}, which is conceptually similar to the widely used generalised Lomb-Scargle periodogram \citep[GLS;][]{Zechmeister2009}, but is better suited for the analysis of multi-telescope data, or datasets with known systematics such as the pre- and post-HARPS fibre intervention data. 
The MLP allows for multiple data subsets, each with its own additive RV offset and jitter term \citep{Zechmeister2019}, while it optimises the log-likelihood $\ln \mathcal{L}$ at each test frequency in a period search.

\begin{figure*}[tp]
    \centering
   \includegraphics[width=9cm,height=15cm]{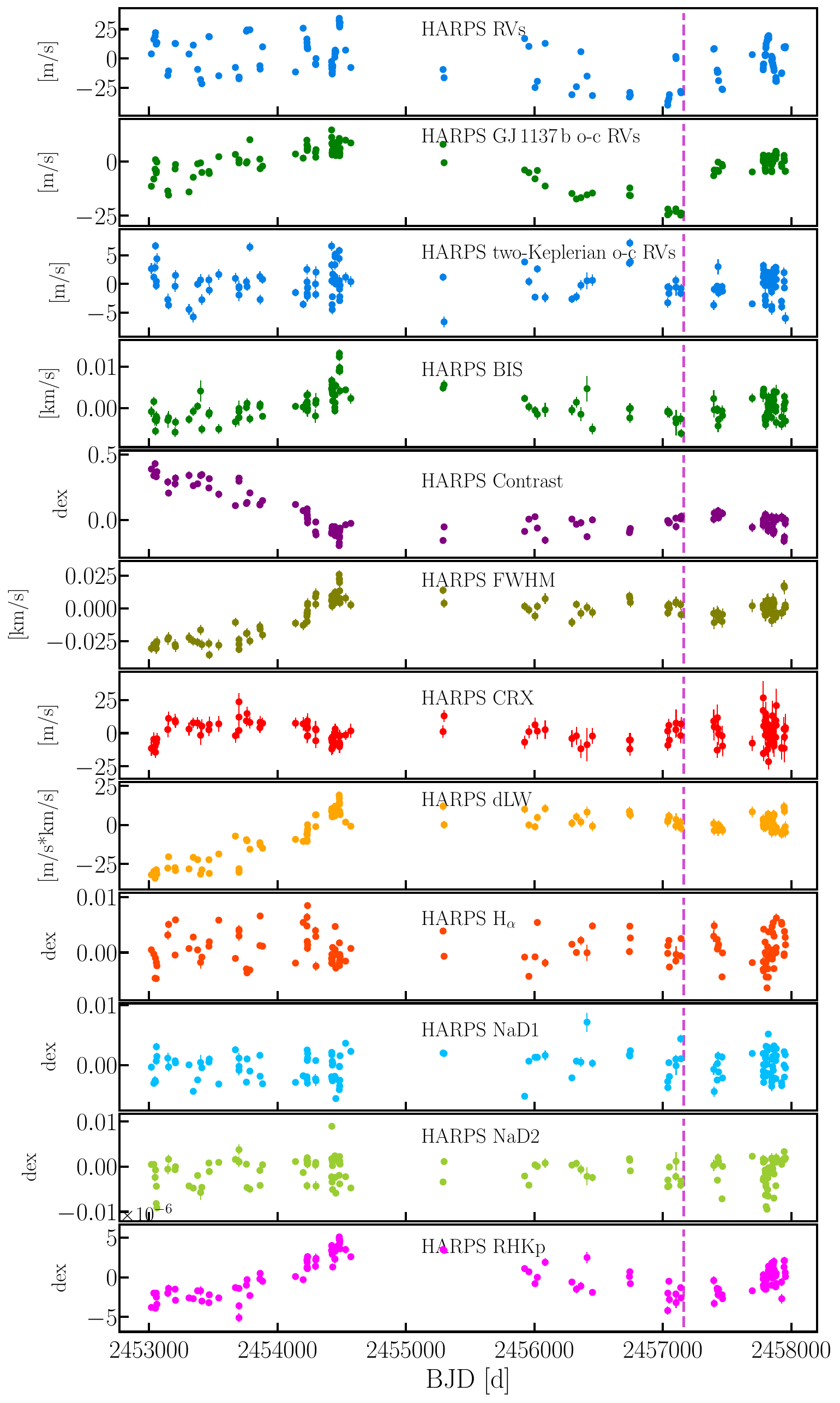} 
   \includegraphics[width=9cm,height=15cm]{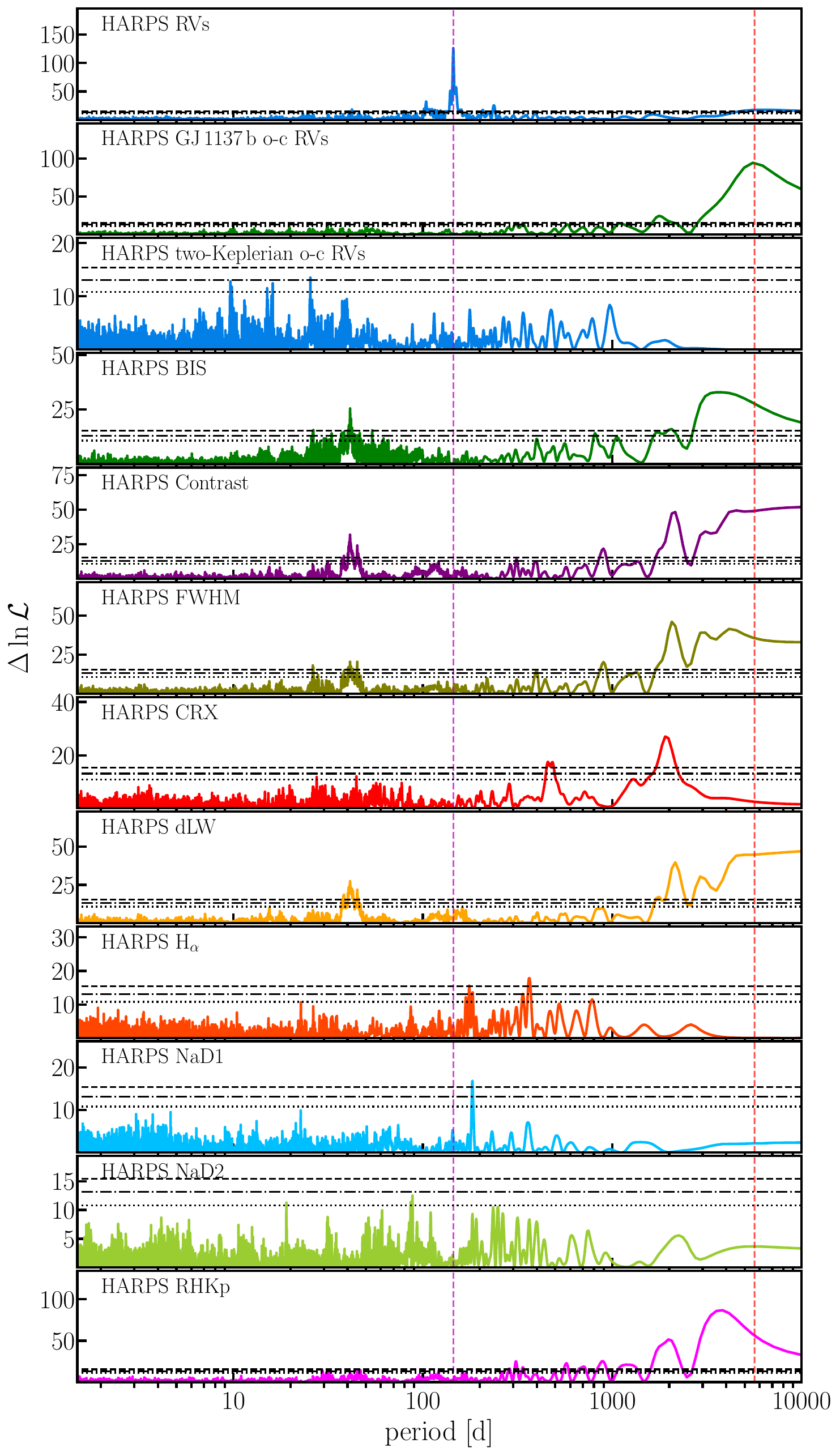} 
    \caption{Time series data and MLS power spectrum of the HARPS RVs and stellar activity indicators of GJ\,1137. The left sub-panels show the individual RVs, the residuals of fitting the GJ\,1137\,b signal, and 
    the two-Keplerian fit residuals including the long-period RV component. The rest are the HARPS activity indices as labelled in the panels. The vertical magenta line notates the HARPS-fibre change from which point we treat the pre- and post-HARPS data with a separate offset parameter. 
    The right panel shows the individual MLP power spectrum in terms of $\Delta lnL$ as indicated in the sub-panels.
    The horizontal lines in the MLP periodograms show the FAP levels of 10\%, 1\%, and 0.1\%, the latter of which is considered
    significant. The magenta vertical line indicates the orbital period of GJ\,1137\,b, whereas the red line indicates the best-fit 
    long-term period seen in the one-planet Keplerian model residuals. 
      }
    \label{MLP_results} 
\end{figure*}

\autoref{MLP_results} presents the results of our MLP analysis of the extracted HARPS spectroscopic data, with individual diagnostics shown in the sub-panels as labelled. The horizontal lines mark the adopted false alarm probability (FAP) thresholds of 10\%, 1\%, and 0.1\%; we consider signals exceeding the 0.1\% threshold significant. Vertical dashed lines indicate the best-fit orbital periods of GJ\,1137\,b and the long-period signal, which we initially suspected to originate from a massive exoplanet with orbital and physical properties comparable to those of Jupiter.

Several HARPS activity indicators, including: BIS, contrast, FWHM, dLW, and $\log R'_{\rm HK}$, however, exhibit significant low-frequency periodicities that closely resemble those found in the RV residuals after modelling the GJ\,1137\,b signal. Given our local example, the Solar System, both Jupiter and the Sun's 11-year magnetic activity cycle induce similar RV signals if the Sun were observed as a star. Therefore, it is not unlikely that GJ\,1137 hosts a massive planet and/or exhibits long-term stellar magnetic cycles that we could detect in spectroscopic datasets. The disentanglement of these phenomena in RV is often challenging and raises caution regarding a possible stellar activity origin for the long-period trend.

\begin{figure*}
    \centering
    \resizebox{\hsize}{!}{\includegraphics{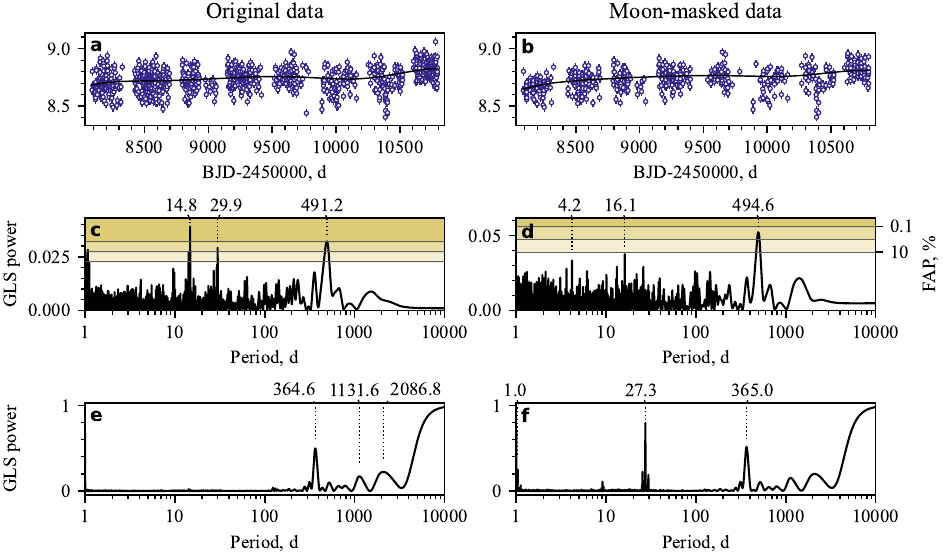}}
    \caption{
    ASAS-SN photometry, periodograms, and window functions. Panels (a) and (b) show the full and Moon-masked photometric datasets, respectively, together with the spline fits used to remove long-term trends. Panels (c) and (d) present the corresponding GLS periodograms, while panels (e) and (f) display their window functions. Our data curation alters the window function, as observations discarded due to Moon contamination occur at regular intervals. In the full dataset, strong peaks appear near $\sim29.9$ and $\sim14.8$ days, corresponding to the lunar synodic period and its first harmonic. These peaks disappear completely in the Moon-masked data. A peak at 491.2 days (full dataset) and 494.6 days (Moon-masked dataset) lies close to the 0.1\% FAP level. However, this signal does not persist coherently across multiple observing windows, and given the overall data quality, we do not consider it further in our analysis.}
    
    \label{fig:assasn_timeseries}
\end{figure*}

\begin{sidewaystable*}

\resizebox{0.72\textheight}{!}
{\begin{minipage}{1.1\textwidth}

\caption{\label{tab:RVs} RVs and activity indices data for GJ\,1137.}
\centering
\begin{tabular}{@{}crrrrrrrrrrrrrrrrrrrrrrr@{}}

\topline
Epoch & RV\tablefootmark{a}  & e\_RV\tablefootmark{a}  & RV\tablefootmark{b}  & e\_RV\tablefootmark{b}  & H$_\alpha$ & e\_H$_\alpha$ & NaD1 & e\_NaD1 & NaD2 & e\_NaD2 & CRX & e\_CRX & dLw & e\_dLw & RHKp & e\_RHKp & FWHM & e\_FWHM & Contrast & e\_Contrast & BIS & e\_BIS & ID set\\                
 BJD & [m/s] & [m/s] & [m/s] & [m/s] &   &   &   &  &  &   & [m/s] & [m/s] & [$m^2/s^2$]   & [$m^2/s^2$] & & & [m/s] & [m/s] &   &  & [m/s] & [m/s] &     \\

\midline

2453017.84490 & 5.33 & 0.95 & 43593.8 & 0.46 & 0.51667 & 0.00059 & 0.1928 & 0.0006 & 0.256 & 0.0008 & -16.16 & 5.45 & -38.011 & 1.143 & 7.3e-06 & 4e-07 & 5913.89 & 3.35 & 53775.9 & 23.52 & -3.74 & 1.28 & 1 \\
2453036.78230 & 17.69 & 0.7 & 43605.9 & 0.4 & 0.5159 & 0.00051 & 0.1901 & 0.0005 & 0.256 & 0.0007 & -9.29 & 5.32 & -35.827 & 1.139 & 9.1e-06 & 4e-07 & 5916.91 & 3.82 & 53726.2 & 26.76 & -1.33 & 1.13 & 1 \\
2453046.66890 & 20.4 & 0.86 & 43609.2 & 0.51 & 0.51155 & 0.00064 & 0.1906 & 0.0007 & 0.2531 & 0.0009 & -14.59 & 5.34 & -40.15 & 1.282 & 9.1e-06 & 4e-07 & 5916.97 & 4.31 & 53817.2 & 30.22 & -5.73 & 1.44 & 1 \\
2453049.71020 & 23.42 & 0.72 & 43611 & 0.37 & 0.51517 & 0.00049 & 0.1904 & 0.0005 & 0.2548 & 0.0006 & -19.03 & 4.51 & -35.246 & 1.146 & 7.4e-06 & 4e-07 & 5916.08 & 3.86 & 53740.2 & 27.02 & -8.45 & 1.07 & 1 \\
2453055.74420 & 13.69 & 0.7 & 43602.7 & 0.42 & 0.51149 & 0.00056 & 0.1938 & 0.0006 & 0.2473 & 0.0007 & -11.8 & 4.59 & -34.74 & 1.24 & 9e-06 & 4e-07 & 5913.73 & 3.91 & 53717.9 & 27.36 & -4.9 & 1.23 & 1 \\
2453056.75370 & 13.82 & 0.94 & 43603.2 & 0.47 & 0.51436 & 0.00059 & 0.1962 & 0.0006 & 0.2511 & 0.0008 & -11.56 & 4.79 & -35.011 & 1.271 & 8.7e-06 & 4e-07 & 5917.57 & 3.96 & 53749.9 & 27.72 & -5.97 & 1.3 & 1 \\
2453060.75310 & 15 & 0.94 & 43603.6 & 0.45 & 0.51379 & 0.00058 & 0.1946 & 0.0006 & 0.2464 & 0.0008 & -8.87 & 4.54 & -37.471 & 1.154 & 7.8e-06 & 4e-07 & 5919.06 & 3.94 & 53756.1 & 27.59 & -5.33 & 1.26 & 1 \\
2453145.55160 & -13.02 & 1.02 & 43574.6 & 0.66 & 0.51935 & 0.00084 & 0.1943 & 0.0009 & 0.2549 & 0.0011 & -1.92 & 5.81 & -33.695 & 1.692 & 9.2e-06 & 5e-07 & 5920.81 & 4.5 & 53677.5 & 31.47 & -5.85 & 1.84 & 1 \\
2453151.57390 & -9.16 & 0.79 & 43575.9 & 0.53 & 0.52126 & 0.00069 & 0.1928 & 0.0007 & 0.2571 & 0.001 & 6.45 & 5.18 & -26.421 & 1.682 & 9.6e-06 & 4e-07 & 5922.14 & 4.11 & 53592.4 & 28.68 & -5.2 & 1.58 & 1 \\
2453202.47620 & 14.24 & 0.96 & 43599.8 & 0.42 & 0.51574 & 0.00054 & 0.1939 & 0.0006 & 0.2548 & 0.0007 & 4.86 & 5.11 & -33.411 & 1.571 & 9.7e-06 & 4e-07 & 5916.77 & 4.2 & 53662.9 & 29.35 & -8.71 & 1.22 & 1 \\
2453205.45340 & 14.05 & 0.94 & 43599.5 & 0.39 & 0.52208 & 0.0005 & 0.1935 & 0.0005 & 0.2555 & 0.0007 & 3.71 & 4.64 & -35.267 & 1.357 & 8.3e-06 & 4e-07 & 5915.23 & 4.23 & 53707.2 & 29.58 & -6.29 & 1.14 & 1 \\
2453310.85550 & 5.21 & 0.93 & 43592.2 & 0.35 & 0.51691 & 0.00045 & 0.1932 & 0.0005 & 0.2512 & 0.0006 & -1.59 & 5 & -34.234 & 1.249 & 8.8e-06 & 3e-07 & 5922.12 & 4.01 & 53728.2 & 28.07 & -5.62 & 1.02 & 1 \\
2453343.85440 & 12.79 & 0.94 & 43600 & 0.39 & 0.51898 & 0.0005 & 0.1888 & 0.0005 & 0.2507 & 0.0007 & 3.15 & 3.89 & -26.796 & 1.309 & 8.5e-06 & 4e-07 & 5919.44 & 3.55 & 53649 & 24.81 & -3.69 & 1.13 & 1 \\
2453377.84190 & -7.84 & 0.7 & 43580.8 & 0.33 & 0.51666 & 0.00043 & 0.1907 & 0.0004 & 0.2535 & 0.0006 & 2.97 & 4.39 & -28.287 & 1.317 & 9.5e-06 & 4e-07 & 5918.63 & 3.37 & 53663.4 & 23.55 & -2.45 & 0.95 & 1 \\
2453400.79650 & -16.54 & 1.06 & 43571.5 & 0.9 & 0.51441 & 0.00117 & 0.1936 & 0.0013 & 0.2497 & 0.0016 & -6.28 & 7.33 & -37.542 & 2.056 & 9.7e-06 & 6e-07 & 5927.93 & 4.05 & 53727.2 & 28.26 & 1.2 & 2.57 & 1 \\
2453410.80630 & -20.04 & 0.94 & 43567.3 & 0.4 & 0.51535 & 0.00052 & 0.1927 & 0.0005 & 0.2509 & 0.0007 & 0.63 & 5.12 & -34.777 & 1.266 & 8.1e-06 & 3e-07 & 5916.78 & 3.93 & 53733.4 & 27.52 & -8 & 1.14 & 1 \\
2453465.67690 & 20.08 & 0.93 & 43607 & 0.38 & 0.51733 & 0.00049 & 0.1937 & 0.0005 & 0.2563 & 0.0007 & -2.08 & 4.9 & -28.3 & 1.377 & 7.9e-06 & 4e-07 & 5917.59 & 3.82 & 53631.9 & 26.66 & -4.4 & 1.1 & 1 \\
2453468.54110 & 19.9 & 0.86 & 43606.8 & 0.47 & 0.51813 & 0.0006 & 0.1941 & 0.0006 & 0.2544 & 0.0008 & 1.99 & 6.01 & -37.078 & 1.614 & 9.1e-06 & 3e-07 & 5908.92 & 3.22 & 53702.7 & 22.53 & -4.01 & 1.33 & 1 \\
2453542.52150 & -13.29 & 0.95 & 43573.8 & 0.43 & 0.52204 & 0.00054 & 0.19 & 0.0006 & 0.2564 & 0.0007 & 2.35 & 5.82 & -24.655 & 1.581 & 8.7e-06 & 4e-07 & 5916.28 & 4.21 & 53583.8 & 29.42 & -7.96 & 1.21 & 1 \\
2453671.86880 & -6.16 & 0.97 & 43580.5 & 0.46 & 0.51513 & 0.00057 & 0.1957 & 0.0006 & 0.2571 & 0.0008 & -6.58 & 5.46 & -13.187 & 1.706 & 1e-05 & 4e-07 & 5933.81 & 3.45 & 53496.4 & 23.98 & -6.23 & 1.29 & 1 \\
\dots \\
\dots \\
\dots \\
2457850.67020 & 0.52 & 0.73 & 43592.6 & 0.35 & 0.52359 & 0.00051 & 0.1909 & 0.0005 & 0.2536 & 0.0006 & 4.08 & 7.34 & 14.569 & 3.053 & 1.02e-05 & 4e-07 & 5958.88 & 4.82 & 53132.8 & 33.03 & 13.88 & 1.05 & 2 \\
2457851.73450 & 1.22 & 0.66 & 43592 & 0.39 & 0.51774 & 0.00056 & 0.1944 & 0.0005 & 0.2507 & 0.0007 & 9.25 & 8.21 & 14.794 & 3.273 & 9.9e-06 & 4e-07 & 5964.64 & 4.81 & 53135.6 & 32.92 & 12.58 & 1.18 & 2 \\
2457852.67810 & 0.31 & 0.63 & 43594.1 & 0.42 & 0.51938 & 0.00061 & 0.1913 & 0.0006 & 0.2538 & 0.0007 & 22.38 & 11.57 & 16.36 & 3.059 & 9.9e-06 & 4e-07 & 5962.21 & 5 & 53121.4 & 34.25 & 13.68 & 1.26 & 2 \\
2457854.74410 & -3.45 & 0.69 & 43585.5 & 0.42 & 0.51836 & 0.00061 & 0.1931 & 0.0006 & 0.2518 & 0.0007 & 1.55 & 6.58 & 18.616 & 3.211 & 1.01e-05 & 6e-07 & 5963.37 & 5.17 & 53088 & 35.41 & 14.57 & 1.28 & 2 \\
2457855.70130 & -4.06 & 0.68 & 43586.6 & 0.36 & 0.52246 & 0.00053 & 0.1948 & 0.0005 & 0.2515 & 0.0006 & 7.04 & 6.62 & 17.784 & 3.129 & 1.09e-05 & 4e-07 & 5965.81 & 5 & 53093.6 & 34.21 & 12.72 & 1.1 & 2 \\
2457858.49470 & -5.52 & 0.64 & 43587.6 & 0.34 & 0.52431 & 0.0005 & 0.1945 & 0.0005 & 0.2535 & 0.0006 & 4.86 & 10.73 & 19.702 & 2.895 & 1.16e-05 & 4e-07 & 5968.23 & 4.98 & 53076.6 & 34.04 & 15.58 & 1.01 & 2 \\
2457859.63560 & -6.21 & 1.25 & 43587.6 & 0.46 & 0.51898 & 0.00067 & 0.1924 & 0.0006 & 0.2531 & 0.0008 & 10.88 & 10.04 & 19.691 & 3.183 & 1.14e-05 & 6e-07 & 5967.68 & 5.04 & 53066.7 & 34.46 & 13.61 & 1.38 & 2 \\
2457864.62810 & -8.05 & 0.81 & 43585.3 & 0.4 & 0.51601 & 0.00058 & 0.1959 & 0.0005 & 0.2451 & 0.0007 & 2.08 & 10.13 & 19.152 & 2.981 & 1.19e-05 & 4e-07 & 5965.9 & 4.77 & 53088.8 & 32.61 & 14.52 & 1.18 & 2 \\
2457865.60490 & -9.77 & 0.65 & 43584.4 & 0.35 & 0.52187 & 0.00051 & 0.1965 & 0.0005 & 0.2505 & 0.0006 & -0.39 & 8.45 & 19.964 & 2.903 & 1.09e-05 & 4e-07 & 5965.24 & 5.2 & 53082.2 & 35.55 & 16.59 & 1.02 & 2 \\
2457878.50650 & -18.61 & 0.69 & 43573.3 & 0.32 & 0.51898 & 0.00047 & 0.1941 & 0.0004 & 0.2531 & 0.0006 & 6.19 & 7.82 & 11.016 & 3.017 & 8.7e-06 & 4e-07 & 5954.6 & 4.52 & 53186.7 & 31.07 & 11.87 & 0.95 & 2 \\
2457879.61460 & -15.63 & 0.62 & 43574.1 & 0.41 & 0.51889 & 0.0006 & 0.191 & 0.0006 & 0.2479 & 0.0007 & 18.77 & 9.88 & 8.921 & 3.258 & 1.01e-05 & 4e-07 & 5955.32 & 5.27 & 53180.2 & 36.2 & 12.9 & 1.27 & 2 \\
2457882.47700 & -17.25 & 0.8 & 43573.7 & 0.47 & 0.51979 & 0.00067 & 0.1961 & 0.0006 & 0.2531 & 0.0008 & 15.21 & 9.1 & 10.263 & 3.19 & 1.03e-05 & 4e-07 & 5957.67 & 4.93 & 53190.2 & 33.83 & 9.13 & 1.37 & 2 \\
2457883.56800 & -18.93 & 0.76 & 43575.2 & 0.59 & 0.52515 & 0.00085 & 0.1949 & 0.0008 & 0.2541 & 0.0011 & 30.11 & 12.73 & 11.587 & 3.491 & 9.2e-06 & 6e-07 & 5958.05 & 5.52 & 53171.2 & 37.86 & 16.47 & 1.76 & 2 \\
2457923.54380 & -12.28 & 0.65 & 43580.4 & 0.45 & 0.52437 & 0.00065 & 0.1904 & 0.0006 & 0.2534 & 0.0008 & -1.64 & 7.92 & 8.856 & 3.165 & 9.1e-06 & 4e-07 & 5958.23 & 4.65 & 53184.2 & 31.93 & 8.64 & 1.35 & 2 \\
2457925.55420 & -11.66 & 0.72 & 43578.4 & 0.44 & 0.52404 & 0.00064 & 0.1899 & 0.0006 & 0.2526 & 0.0008 & -2.2 & 8.61 & 7.292 & 2.998 & 7.5e-06 & 5e-07 & 5957.74 & 4.78 & 53194.3 & 32.85 & 10.47 & 1.34 & 2 \\
2457943.50200 & 10.4 & 0.87 & 43603.7 & 0.42 & 0.51838 & 0.00061 & 0.1933 & 0.0006 & 0.2539 & 0.0008 & 12.48 & 8.79 & 24.912 & 3.268 & 1.12e-05 & 5e-07 & 5978.81 & 4.5 & 53009 & 30.65 & 12.39 & 1.32 & 2 \\
2457944.50320 & 9.76 & 0.72 & 43603.1 & 0.39 & 0.51867 & 0.00056 & 0.1955 & 0.0006 & 0.2554 & 0.0007 & -2.15 & 10.51 & 24.168 & 3.424 & 1.2e-05 & 6e-07 & 5978.45 & 4.49 & 53018.1 & 30.57 & 13.35 & 1.22 & 2 \\
2457946.46070 & 10.3 & 0.56 & 43603.1 & 0.33 & 0.52272 & 0.00048 & 0.1913 & 0.0005 & 0.2536 & 0.0006 & 6.37 & 10.46 & 21.934 & 3.031 & 1.19e-05 & 5e-07 & 5977.62 & 5.21 & 53040.4 & 35.51 & 15.38 & 1.02 & 2 \\
2457952.46600 & 9.56 & 1.05 & 43602.3 & 0.47 & 0.52167 & 0.00068 & 0.195 & 0.0007 & 0.2539 & 0.0008 & 12.06 & 7.43 & 12.066 & 3.359 & 1.03e-05 & 5e-07 & 5964.1 & 5.02 & 53142.8 & 34.38 & 14.06 & 1.42 & 2 \\
2457954.47040 & 10.74 & 1.04 & 43601.6 & 0.47 & 0.52032 & 0.00068 & 0.1929 & 0.0007 & 0.254 & 0.0009 & 13.39 & 10.42 & 8.539 & 3.335 & 1.01e-05 & 5e-07 & 5962.06 & 5.46 & 53171.8 & 37.47 & 9.5 & 1.46 & 2 \\
\bottomline
\end{tabular}
\tablefoot{
\tablefoottext{a}{Differential RV from SERVAL obtained from the {\sc HARPS-RVBank} \citep{Trifonov2020,Perdelwitz2024}}.
\tablefoottext{b}{Absolute CCF RV from RACOON \citep{Lafarga2020}}.
{\bf Note:} The H$_\alpha$, NaD1, NaD2, CRX, dLw, and RHKp measurements are taken from the {\sc HARPS-RVBank}, whereas FWHM, Contrast, and BIS have been derived with RACOON. ID set 1 notates the HARPS spectra pre-fibre upgrade in May 2015 \citep{LoCurto2015}, whereas ID set 2 notates the post-fibre upgrade data.\\  

} 
\end{minipage}}
\end{sidewaystable*}

\begin{table*}
    \centering
    \caption{
        Parameter posteriors of our best model that included only an LTF and GJ~1137~b.
    }
    \label{tab:model_cycle_posterior}
    
    \begin{tabular}{lcccc}
    \hline \hline
    Parameter name &
    Symbol &
    Unit &
    Prior &
    Posterior \\\hline
    \textbf{LTF parameters} \\
    Period &
    $P_\text{cyc}$ &
    \unit\day &
    $\mathcal{U}_{\log}\left(10^3, 10^4\right)$ &
    $5320^{+170}_{-150}$ \\
    RV phase &
    $\varphi_\text{cyc, 0}$ &
    - &
    $\mathcal{U}\left(0, 1\right)$\tablefootmark{w} &
    $0.894\pm 0.020$ \\
    RV semi-amplitude &
    $k_\text{cyc, 0}$ &
    \unit{\metre\per\second} &
    $\mathcal{U}\left(0, 40\right)$ &
    $14.2\pm 0.6$ \\
    FWHM phase &
    $\varphi_\text{cyc, 1}$ &
    - &
    $\mathcal{U}\left(0, 1\right)$\tablefootmark{w} &
    $0.888^{+0.021}_{-0.022}$ \\
    FWHM semi-amplitude &
    $k_\text{cyc, 1}$ &
    \unit{\metre\per\second} &
    $\mathcal{U}\left(0, 40\right)$ &
    $18.7^{+1.8}_{-1.7}$ \\
    $R'_\text{HK}$ phase &
    $\varphi_\text{cyc, 2}$ &
    - &
    $\mathcal{U}\left(0, 1\right)$\tablefootmark{w} &
    $0.797^{+0.021}_{-0.020}$ \\
    $R'_\text{HK}$ semi-amplitude &
    $k_\text{cyc, 2}$ &
    \unit\ppm &
    $\mathcal{U}\left(0, 10\right)$ &
    $6.59^{+0.89}_{-0.79}$ \\
    RV zero-order correction &
    $\alpha_{0,0}$ &
    \unit{\metre\per\second} &
    $\mathcal{N}(\mu_\text{lm},200\sigma_\text{lm})$ &
    $-3.64^{+0.57}_{-0.60}$ \\
    FWHM zero-order correction &
    $\alpha_{0,1}$ &
    \unit{\metre\per\second} &
    $\mathcal{N}(\mu_\text{lm},200\sigma_\text{lm})$ &
    $-5.86^{+1.53}_{-1.57}$ \\
    $R'_\text{HK}$ second-order correction &
    $\alpha_{2,2}$ &
    \unit{\ppm\per\day\squared} &
    $\mathcal{N}(\mu_\text{lm},200\sigma_\text{lm})$ &
    $(1.10^{+0.28}_{-0.27})\times 10^{-6}$ \\
    $R'_\text{HK}$ first-order correction &
    $\alpha_{1,2}$ &
    \unit{\ppm\per\day} &
    $\mathcal{N}(\mu_\text{lm},200\sigma_\text{lm})$ &
    $(6.60^{+1.49}_{-1.41})\times 10^{-3}$ \\
    $R'_\text{HK}$ zero-order correction &
    $\alpha_{0,2}$ &
    \unit{\ppm} &
    $\mathcal{N}(\mu_\text{lm},200\sigma_\text{lm})$ &
    $7.19^{+1.33}_{-1.31}$ \\
    HARPS-pre FWHM linear drift &
    $\beta$ &
    \unit{\metre\per\second\per\day} &
    $\mathcal{U}(-0.1,0.1)$ &
    $0.0127^{+(17)}_{-(16)}$ \\
    \textbf{Dataset parameters} \\
    HARPS-post RV offset &
    $O_{1,0}$ &
    \unit{\metre\per\second} &
    $\mathcal{N}(0,5\sigma_0)$ &
    $9.78^{+1.84}_{-1.86}$ \\
    HARPS-post FWHM offset &
    $O_{1,1}$ &
    \unit{\metre\per\second} &
    $\mathcal{N}(0,5\sigma_1)$ &
    $20.1^{+3.0}_{-2.8}$ \\
    HARPS-post $R'_\text{HK}$ offset &
    $O_{1,2}$ &
    \unit{\ppm} &
    $\mathcal{N}(0,5\sigma_2)$ &
    $0.35\pm 0.71$ \\
    HARPS-pre RV jitter &
    $J_{0,0}$ &
    \unit{\metre\per\second} &
    $\mathcal{U}_{\log}\left(10^{-3}, 10^3\right)$ &
    $3.03^{+0.29}_{-0.24}$ \\
    HARPS-post RV jitter &
    $J_{1,0}$ &
    \unit{\metre\per\second} &
    $\mathcal{U}_{\log}\left(10^{-3}, 10^3\right)$ &
    $2.17^{+0.28}_{-0.24}$ \\
    HARPS-pre FWHM jitter &
    $J_{0,1}$ &
    \unit{\metre\per\second} &
    $\mathcal{U}_{\log}\left(10^{-3}, 10^3\right)$ &
    $6.94^{+0.73}_{-0.66}$ \\
    HARPS-post FWHM jitter &
    $J_{1,1}$ &
    \unit{\metre\per\second} &
    $\mathcal{U}_{\log}\left(10^{-3}, 10^3\right)$ &
    $0.22^{+2.26}_{-0.22}$ \\
    HARPS-pre $R'_\text{HK}$ jitter &
    $J_{0,2}$ &
    \unit{\ppm} &
    $\mathcal{U}_{\log}\left(10^{-2}, 10^3\right)$ &
    $1.12^{+0.11}_{-0.10}$ \\
    HARPS-post $R'_\text{HK}$ jitter &
    $J_{1,2}$ &
    \unit{\ppm} &
    $\mathcal{U}_{\log}\left(10^{-2}, 10^3\right)$ &
    $0.90^{+0.12}_{-0.11}$ \\
    \textbf{GJ~1137~b} \\
    Period &
    $P_\text{b}$ &
    \unit{\day} &
    $\mathcal{U}\left(140, 150\right)$ &
    $144.751\pm 0.030$ \\
    Phase &
    $\varphi_\text{b}$ &
    - &
    $\mathcal{U}\left(0, 1\right)$\tablefootmark{w} &
    $0.704^{+0.07}_{-0.06}$ \\
    RV semi-amplitude &
    $k_\text{rv, b}$ &
    \unit{\metre\per\second} &
    $\mathcal{U}\left(0, 30\right)$ &
    $20.0\pm 0.4$ \\
    Eccentricity &
    $e_\text{b}$ &
    - &
    $\mathcal{U}\left(0, 1\right)$ &
    $0.105^{+0.016}_{-0.017}$ \\
    Argument of periastron &
    $\omega_\text{b}$ &
    - &
    $\mathcal{U}\left(0, 2\pi\right)$\tablefootmark{w} &
    $6.08^{+0.20}_{-0.18}$ \\
    \end{tabular}
    \tablefoot{
    \tablefoottext{w}{Wrapped parameter.}
    Reported uncertainties reflect the 16\textsuperscript{th} and the 84\textsuperscript{th} percentiles. The standard deviation of measurements in physical quantity $j$ is denoted $\sigma_j$.
    }
\end{table*}

\begin{figure*}
    \centering
    \resizebox{\hsize}{!}{\includegraphics{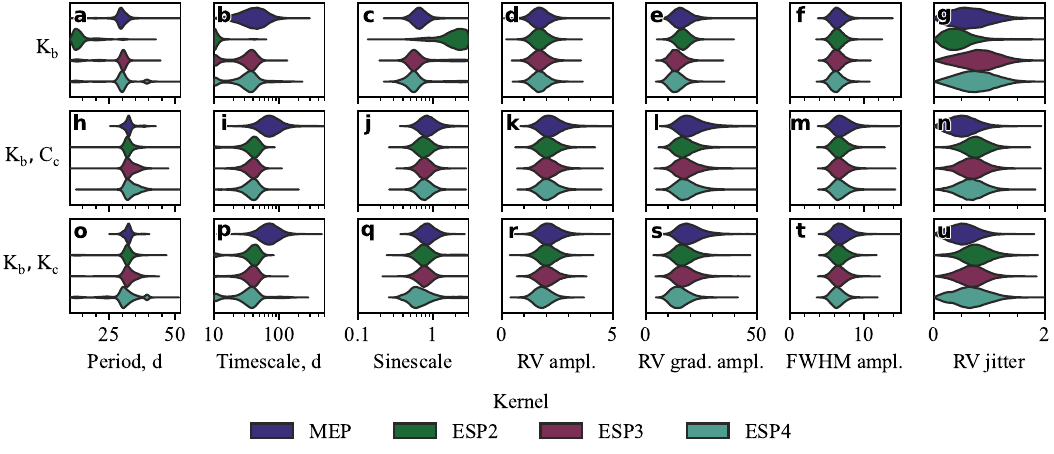}}
    \caption{
    Violin plots of models in our model grid (Fig.~\ref{fig:model_table}), that show the posteriors of the GP hyperparameters as well as the RV jitter.
    }
    \label{fig:kernel_comparison}
\end{figure*}

\begin{table}
\small
\centering
\caption{
Period posteriors of GJ~1137~c in units of \unit{\day} from the grid-search in Fig.~\ref{fig:model_table}.
}
\label{tab:model_comparison_planetc_period}

\begin{tabular}{ccccc}
\hline\hline
Planetary & \multicolumn{4}{c}{Stellar-activity kernel} \\ \cline{2-5}
config. & MEP & ESP2 & ESP3 & ESP4 \\ \hline
K$_\text{b}$C$_\text{c}$ & $9.6412^{+(12)}_{-(11)}$&$9.6412(11)$&$9.6411^{+(12)}_{-(11)}$&$9.6412(11)$ \\
K$_\text{b}$K$_\text{c}$ & $9.6412^{+(13)}_{-(12)}$&$9.6411^{+(14)}_{-(12)}$&$9.6412^{+(16)}_{-(13)}$&$9.64^{+22.69}_{-2.71}$ \\ \hline
\end{tabular}
\tablefoot{
Reported uncertainties reflect the 16\textsuperscript{th} and the 84\textsuperscript{th} percentiles. Uncertainties are given in parentheses, to the order of the least significant figure.
}
\end{table}

\begin{table}
\small
\centering
\caption{
Phase posteriors of GJ~1137~c from the grid-search in Fig.~\ref{fig:model_table}.
}
\label{tab:model_comparison_planetc_phase}

\begin{tabular}{ccccc}
\hline\hline
Planetary & \multicolumn{4}{c}{Stellar-activity kernel} \\ \cline{2-5}
config. & MEP & ESP2 & ESP3 & ESP4 \\ \hline
K$_\text{b}$C$_\text{c}$ & $0.310\pm 0.028$&$0.310^{+0.026}_{-0.027}$&$0.315^{+0.029}_{-0.030}$&$0.315^{+0.029}_{-0.031}$ \\
K$_\text{b}$K$_\text{c}$ & $0.303^{+0.046}_{-0.048}$&$0.292^{+0.052}_{-0.056}$&$0.301^{+0.056}_{-0.057}$&$0.345^{+0.259}_{-0.109}$ \\ \hline
\end{tabular}
\tablefoot{
Reported uncertainties reflect the 16\textsuperscript{th} and the 84\textsuperscript{th} percentiles.
}
\end{table}

\begin{table}
\small
\centering
\caption{
RV semi-amplitude posteriors of GJ~1137~c in units of \unit{\metre\per\second} from the grid-search in Fig.~\ref{fig:model_table}.
}
\label{tab:model_comparison_planetc_krv}

\begin{tabular}{ccccc}
\hline\hline
Planetary & \multicolumn{4}{c}{Stellar-activity kernel} \\ \cline{2-5}
config. & MEP & ESP2 & ESP3 & ESP4 \\ \hline
K$_\text{b}$C$_\text{c}$ & $1.73^{+0.24}_{-0.23}$&$1.67\pm 0.23$&$1.68^{+0.24}_{-0.25}$&$1.70^{+0.24}_{-0.25}$ \\
K$_\text{b}$K$_\text{c}$ & $1.71^{+0.23}_{-0.24}$&$1.65^{+0.23}_{-0.26}$&$1.65^{+0.25}_{-0.30}$&$1.35^{+0.47}_{-0.71}$ \\ \hline
\end{tabular}
\tablefoot{
Reported uncertainties reflect the 16\textsuperscript{th} and the 84\textsuperscript{th} percentiles.
}
\end{table}

\begin{table}
\small
\centering
\caption{
Eccentricity posteriors of GJ~1137~c from the grid-search in Fig.~\ref{fig:model_table}.
}
\label{tab:model_comparison_planetc_eccentricity}

\begin{tabular}{ccccc}
\hline\hline
Planetary & \multicolumn{4}{c}{Stellar-activity kernel} \\ \cline{2-5}
config. & MEP & ESP2 & ESP3 & ESP4 \\ \hline
K$_\text{b}$C$_\text{c}$ & -&-&-&- \\
K$_\text{b}$K$_\text{c}$ & $0.10^{+0.10}_{-0.07}$&$0.15^{+0.12}_{-0.10}$&$0.13^{+0.13}_{-0.09}$&$0.21^{+0.53}_{-0.14}$ \\ \hline
\end{tabular}
\tablefoot{
Reported uncertainties reflect the 16\textsuperscript{th} and the 84\textsuperscript{th} percentiles.
}
\end{table}

\begin{table}
\small
\centering
\caption{
Argument-of-periastron posteriors of GJ~1137~c in units of \unit\radian~from the grid-search in Fig.~\ref{fig:model_table}.
}
\label{tab:model_comparison_planetc_omega}

\begin{tabular}{ccccc}
\hline\hline
Planetary & \multicolumn{4}{c}{Stellar-activity kernel} \\ \cline{2-5}
config. & MEP & ESP2 & ESP3 & ESP4 \\ \hline
K$_\text{b}$C$_\text{c}$ & -&-&-&- \\
K$_\text{b}$K$_\text{c}$ & $4.04^{+1.08}_{-1.79}$&$4.09^{+0.88}_{-1.07}$&$4.08^{+0.96}_{-1.26}$&$4.04^{+1.51}_{-2.02}$ \\ \hline
\end{tabular}
\tablefoot{
Reported uncertainties reflect the 16\textsuperscript{th} and the 84\textsuperscript{th} percentiles.
}
\end{table}

\begin{figure}
    \centering
    \resizebox{\hsize}{!}{\includegraphics{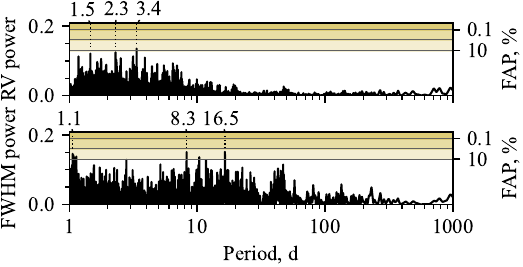}}
    \caption{
    GLSPs of the residual time series of our best model, accounting for the model jitter, for: (a) SERVAL RVs, (b) RACCOON FWHMs.
    }
    \label{fig:bestmodel_pgram}
\end{figure}

\begin{figure}
    \centering
    \resizebox{\hsize}{!}{\includegraphics{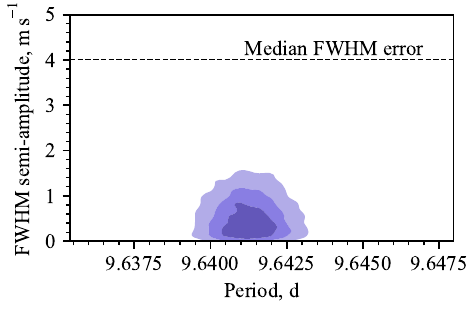}}
    \caption{
    FWHM amplitude-period distribution in our search for a common RV\&FWHM signal in the data. The shaded contours highlight 25\%, 50\%, and 75\% of the enclosed probability mass. The median FWHM error is marked with a solid dashed line.
    }
    \label{fig:pseudoplanet_posterior}
\end{figure}

\begin{figure}
    \centering
    \resizebox{\hsize}{!}{\includegraphics{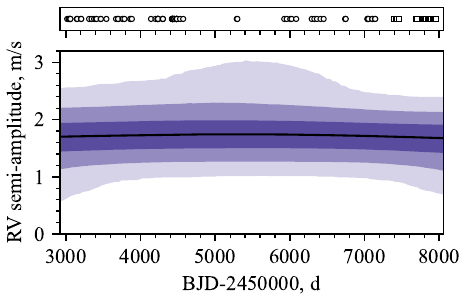}}
    \caption{
    Stability of the RV signal induced by GJ\,1137\,c through an apodisation test. Centre: the median value (solid black line), as well as the $1\sigma$, $2\sigma$, and $3\sigma$ confidence intervals (shaded bands) of the apodised signal over time. Top: timestamps of measurements, with markers following Fig.~\ref{fig:timeseries}.
    }
    \label{fig:apo_signal}
\end{figure}

\begin{table*}
    \centering
    \caption{
        Parameter posteriors of our best model, against the results published by \citet{Lovis2005}.
    }
    \label{tab:bestmodel_posterior}
    
    \begin{tabular}{lcclcc}
    \hline \hline
    Parameter name &
    Symbol &
    Unit &
    Prior &
    \multicolumn{2}{c}{Posterior} \\ \cline{5-6}
     &
     &
     &
     &
    This work &
    \citet{Lovis2005} \\ \hline
    \textbf{LTF parameters} \\
    Period &
    $P_\text{cyc}$ &
    \unit\day &
    $\mathcal{U}_{\log}\left(4000, 10^4\right)$ &
    $5870^{+480}_{-350}$ &
    - \\
    RV phase &
    $\varphi_\text{cyc, 0}$ &
    - &
    $\mathcal{U}\left(0, 1\right)$\tablefootmark{w} &
    $0.837^{+0.037}_{-0.043}$ &
    - \\
    RV semi-amplitude &
    $k_\text{cyc, 0}$ &
    \unit{\metre\per\second} &
    $\mathcal{U}\left(0, 40\right)$ &
    $14.6^{+1.3}_{-1.2}$ &
    - \\
    FWHM phase &
    $\varphi_\text{cyc, 1}$ &
    - &
    $\mathcal{U}\left(0, 1\right)$\tablefootmark{w} &
    $0.864^{+0.054}_{-0.070}$ &
    - \\
    FWHM semi-amplitude &
    $k_\text{cyc, 1}$ &
    \unit{\metre\per\second} &
    $\mathcal{U}\left(0, 40\right)$ &
    $18.4^{+3.5}_{-3.9}$ &
    - \\
    RV zero-order correction &
    $\alpha_{0,0}$ &
    \unit{\metre\per\second} &
    $\mathcal{N}(\mu_\text{lm},200\sigma_\text{lm})$ &
    $-4.76^{+1.18}_{-1.49}$ &
    - \\
    FWHM zero-order correction &
    $\alpha_{0,1}$ &
    \unit{\metre\per\second} &
    $\mathcal{N}(\mu_\text{lm},200\sigma_\text{lm})$ &
    $-9.71^{+2.22}_{-2.27}$ &
    - \\
    HARPS-pre FWHM linear drift &
    $\beta$ &
    \unit{\metre\per\second\per\day} &
    $\mathcal{U}(-0.1,0.1)$ &
    $0.0150^{+(23)}_{-(28)}$ &
    - \\
    \textbf{Dataset parameters} \\
    HARPS-post RV offset &
    $O_{1,0}$ &
    \unit{\metre\per\second} &
    $\mathcal{N}(0,5\sigma_0)$ &
    $13.4^{+3.3}_{-3.2}$ &
    - \\
    HARPS-post FWHM offset &
    $O_{1,1}$ &
    \unit{\metre\per\second} &
    $\mathcal{N}(0,5\sigma_1)$ &
    $23.2^{+6.2}_{-6.4}$ &
    - \\
    HARPS-pre RV jitter &
    $J_{0,0}$ &
    \unit{\metre\per\second} &
    $\mathcal{U}(0, 5)$ &
    $0.63^{+0.38}_{-0.37}$ &
    - \\
    HARPS-post RV jitter &
    $J_{1,0}$ &
    \unit{\metre\per\second} &
    $\mathcal{U}(0, 5)$ &
    $0.29^{+0.26}_{-0.20}$ &
    - \\
    HARPS-pre FWHM jitter &
    $J_{0,1}$ &
    \unit{\metre\per\second} &
    $\mathcal{U}(0, 5)$ &
    $1.97^{+1.09}_{-1.14}$ &
    - \\
    HARPS-post FWHM jitter &
    $J_{1,1}$ &
    \unit{\metre\per\second} &
    $\mathcal{U}(0, 5)$ &
    $0.67^{+0.75}_{-0.48}$ &
    - \\
    \textbf{Stellar-activity parameters} \\
    Timescale &
    $\tau$ &
    \unit\day &
    $\mathcal{U}_{\log}\left(10, 10^{4}\right)$ &
    $73^{+33}_{-21}$ &
    - \\
    Period &
    $P_\text{rot}$ &
    \unit\day &
    $\mathcal{U}\left(10,52\right)$ &
    $32.3^{+1.2}_{-1.3}$ &
    - \\
    Sinescale (harmonic complexity) &
    $\eta$ &
    - &
    $\mathcal{U}_{\log}\left(0.1,3\right)$ &
    $0.85^{+0.26}_{-0.17}$ &
    - \\
    RV amplitude &
    $A_0$ &
    \unit{\metre\per\second} &
    $\mathcal{U}(-10^3,10^3)$ &
    $2.20^{+0.53}_{-0.42}$ &
    - \\
    RV gradient amplitude &
    $B_0$ &
    \unit{\metre\per\second\per\day} &
    $\mathcal{U}(-10^3,10^3)$ &
    $19.9^{+6.9}_{-4.7}$ &
    - \\
    FWHM amplitude &
    $A_1$ &
    \unit{\metre\per\second} &
    $\mathcal{U}_{\log}\left(10^{-3}, 10^{3}\right)$ &
    $6.99^{+1.33}_{-0.99}$ &
    - \\
    \textbf{GJ~1137~b} \\
    Period &
    $P_\text{b}$ &
    \unit{\day} &
    $\mathcal{U}\left(140, 150\right)$ &
    $144.720\pm 0.029$ &
    $143.58\pm 0.60$ \\
    Phase &
    $\varphi_\text{b}$ &
    - &
    $\mathcal{U}\left(0, 1\right)$\tablefootmark{w} &
    $0.706\pm 0.006$ &
    - \\
    Inferior-conjunction ephemeris &
    $\varepsilon_\text{b}$ &
    \unit\julianday$-\num{2450000}$ &
    derived &
    $7852.24^{+0.89}_{-0.85}$ &
    $3181.7\pm 3.0$ \\
    RV semi-amplitude &
    $k_\text{rv, b}$ &
    \unit{\metre\per\second} &
    $\mathcal{U}\left(0, 30\right)$ &
    $19.8\pm 0.4$ &
    $18.3\pm 0.5$ \\
    Eccentricity &
    $e_\text{b}$ &
    - &
    $\mathcal{U}\left(0, 1\right)$ &
    $0.118^{+0.016}_{-0.015}$ &
    $0.14\pm 0.03$ \\
    Argument of periastron &
    $\omega_\text{b}$ &
    \unit\radian &
    $\mathcal{U}\left(0, 2\pi\right)$\tablefootmark{w} &
    $5.83\pm 0.15$ &
    $5.82\pm 0.14$ \\
    Semi-major axis &
    $a_\text{b}$ &
    \unit\au &
    derived &
    $0.508\pm 0.005$ &
    $0.477$ \\
    Minimum mass &
    $m_\text{b}\sin i_\text{b}$ &
    $M_J$ &
    derived &
    $0.451\pm 0.012$ &
    $0.37$ \\
    Temporal-average incident flux &
    $\Phi_\text{b}$ &
    \unit\earthflux &
    derived &
    $1.59\pm 0.15$ &
    - \\
    \textbf{GJ~1137~c} \\
    Period &
    $P_\text{c}$ &
    \unit{\day} &
    $\mathcal{U}_{\log}\left(1, 100\right)$ &
    $9.6412^{+(12)}_{-(11)}$ &
    - \\
    Phase &
    $\varphi_\text{c}$ &
    - &
    $\mathcal{U}\left(0, 1\right)$\tablefootmark{w} &
    $0.310\pm 0.028$ &
    - \\
    Inferior-conjunction ephemeris &
    $\varepsilon_\text{c}$ &
    \unit\julianday$-\num{2450000}$ &
    derived &
    $7951.49\pm 0.27$ &
    - \\
    RV semi-amplitude &
    $k_\text{rv, c}$ &
    \unit{\metre\per\second} &
    $\mathcal{U}\left(0, 10\right)$ &
    $1.73^{+0.24}_{-0.23}$ &
    - \\
    Eccentricity &
    $e_\text{c}$ &
    - &
    - &
    0 &
    - \\
    Semi-major axis &
    $a_\text{c}$ &
    \unit\au &
    derived &
    $0.0835(8)$ &
    - \\
    Minimum mass &
    $m_\text{c}\sin i_\text{c}$ &
    \unit\earthmass &
    derived &
    $5.12^{+0.70}_{-0.69}$ &
    - \\
    Temporal-average incident flux &
    $\Phi_\text{c}$ &
    \unit\earthflux &
    derived &
    $58.4^{+5.6}_{-5.5}$ &
    - \\ \hline
    \end{tabular}
    \tablefoot{
    \tablefoottext{w}{Wrapped parameter.}
    Reported uncertainties reflect the 16\textsuperscript{th} and the 84\textsuperscript{th} percentiles. The standard deviation of measurements in physical quantity $j$ is denoted $\sigma_j$. Uncertainties of $P_\text{c}$ and $a_\text{c}$ are given in parentheses, to the order of the least significant figure.
    }
\end{table*}

\end{appendix}

\end{document}